\documentclass[12pt,preprint]{aastex}
\shorttitle{Rotation in IC 348}
\shortauthors{Nordhagen et al.}

\begin{document}

\title{The Variability and Rotation of Pre-main Sequence Stars  in IC 348: Does Intracluster Environment Influence Stellar Rotation?}

\author{Stella Nordhagen\altaffilmark{1}}
\affil{Department of Physics, Middlebury College, Middlebury, VT 05753}
\email{snordhag@middlebury.edu}

\and

\author{William Herbst, Katherine L. Rhode and Eric C. Williams}
\affil{Astronomy Department, Wesleyan University, Middletown, CT 06459}

\altaffiltext{1}{REU summer student at Wesleyan University}

\begin{abstract}

A variability study of the young cluster IC 348 at Van Vleck
Observatory has been extended to a total of seven years. Twelve new
periodic stars have been found in the last two years, bringing the
total discovered by this program to 40. In addition, we confirm 16 of
the periods reported by others and resolve some discrepancies. The
total number of known rotation periods in the cluster, from all
studies has now reached 70. This is sufficient to demonstrate that the
parent population of K5-M2 stars is rotationally indistinguishable
from that in the Orion Nebula Cluster even though their radii are 20\%
smaller and they would be expected to spin about twice as fast if
angular momentum were conserved. The median radius and, therefore,
inferred age of the IC 348 stars actually closely matches that of NGC
2264, but the stars spin significantly more slowly. This suggests that
another factor besides mass and age plays a role in establishing the
rotation properties within a cluster and we suggest that it is
environment. If disk locking were to persist for longer times in less
harsh environments, because the disks themselves persist for longer
times, it could explain the generally slower rotation rates observed
for stars in this cluster, whose earliest type star is of class B5. We
have also obtained radial velocities, the first for PMS stars in
IC348, and \emph{v} sin \emph{i} measurements for 30 cluster stars to
assist in the study of rotation and as an independent check on stellar
radii. Several unusual variable stars are discussed; in some or all
cases their behavior may be linked to occultations by circumstellar
material.  A strong correlation exists between the range of
photometric variability and the slope of the spectral energy
distribution in the infrared. Nineteen of the 21 stars with I ranges
exceeding 0.4 mag show infrared evidence for circumstellar disks.

\end{abstract}


\section{Introduction}

IC 348 is a compact, young cluster in the Perseus clouds at a
distance of about 300 pc \citep{c93}. It contains $\sim 200$ members, with spectral types ranging from B5 to late
M and about two dozen probable brown dwarfs that are also likely to
be cluster members \citep{l03,llme05}. It is especially important to
star formation studies because it is relatively close to us, compact
on the sky, well studied by a variety of methods, and has a nearly
complete census of members. It is also intermediate in nature
between the denser, more massive clusters, such as the Orion Nebula
Cluster (ONC), and the looser associations, such as the Taurus T
association. For a recent summary of work on the cluster, see the article by \citet{h06} in {\it The Handbook of Low Mass Star Forming Regions}.

Photometric monitoring of young clusters has led to the determination of accurate rotation periods for about 2,000 pre-main sequence stars (see, for example, the review by \citet{hems06}). The best-studied cases are the ONC and NGC 2264. \citet{hm05} have recently reviewed the data and shown that the evolution of rotation for stars in the 0.4-1.2 solar mass range may be understood in terms of two phenomena. About half of the stars of Orion age (the half which are already the more rapid rotators) spin up with essentially no loss of angular momentum all the way to the ZAMS. The other half continue to be slowed by interaction with circumstellar disks for times of up to 5 Myr and end up as very slow rotators on the ZAMS. This picture is based on only five clusters, two containing pre-main sequence stars and three containing ZAMS stars. Obviously it will be important to test it by obtaining data for more clusters and, in particular, clusters with different properties (such as stellar density and environment). IC 348 is a particularly useful cluster in this regard for the reasons discussed in the first paragraph.

The cluster has been photometrically monitored at Van Vleck Observatory (VVO) on the campus of Wesleyan University for seven years. Two previous papers in this series have reported on the progress of this program after one and five years, respectively \citep{hmw00, chw04}.
\citet{lit05} have recently published results from a short (17 day) but intensive photometric monitoring program done with a larger telescope and going deeper than our work, and \citet{kkb05} have even more recently published the results of a six month monitoring program designed to search for variability in stars with X-ray counterparts. In this paper, we present data from an additional two years of photometric monitoring at VVO as well as a complementary study of  radial velocities and \emph{v} sin \emph{i} measurements for 30 cluster members. We combine the results of all of the studies to discuss the rotation period distribution in the cluster and its significance for our understanding of angular momentum evolution in solar-like stars. A very recent infrared study of the cluster with {\it Spitzer} by \citet{lm06} allows us to correlate rotation and variability properties with the presence or absence of evidence for a disk. In addition, we discuss some unusual variable stars.

\section{Observations and Reductions}

\subsection{Photometry}

Observations of IC 348 at Wesleyan have been obtained between 1998
December 10 and 2005 March 17. Earlier papers in this series
\citep{hmw00, chw04} have reported results for the first five years; here we concentrate on the last two years. All data were obtained
using a 1024 x 1024 Photometrics CCD camera attached to the 0.6 m
Perkin telescope at Wesleyan University's Van Vleck Observatory. As
each pixel covers 0.6\arcsec, the field of view is 10.2\arcmin\ on a
side. On each clear night during the observing season, five consecutive
one-minute exposures were taken through a Cousins \emph{I} filter,
along with bias frames, dark frames, and twilight flats. When
possible, this sequence of five object frames was repeated more than once per night.
Preliminary reductions were conducted using standard Image Reduction
and Analysis Facility (IRAF) tasks, and each set of five images was
combined and shifted to the same position within less than one pixel
to create an image with an increased dynamic range and an effective
five-minute exposure time.

Differential aperture photometry was performed on a sample of 151
stars, listed in \citet{chw04}, using the IRAF APPHOT package with
an aperture radius of 7 pixels (4.2\arcsec). The median level of sky
background was determined using an annulus with inner and outer
radii of 10 and 15 pixels, respectively. Due to the close proximity
of other stars, a few stars in the sample may have contaminated
photometry. A list of these objects can be found in \citet{hmw00}.

As the observations were taken over a seven-year interval, it was
beneficial to have a set of stable comparison stars that could be
used reliably over the entire time span. For this reason, the six
comparison stars established by  \citet{chw04} were used. All six
stars (3, 7, 8, 10, 13, and 17) display standard deviations of less
than 0.01 mag in each season. This stability is likely due to the
first five stars' proximity to the main sequence and the sixth
star's non-cluster-member status. Instrumental differential
magnitudes (\emph{i}) were computed for each star on each night
relative to the average magnitude of comparison stars. These values
were transformed to standard magnitude (\emph{I}) using
the adopted transformation
\begin{equation}
I = i + 11.142 + 0.099 (R-I)
\end{equation}
The values of \emph{R-I} used in the transformation are taken from \citet{h98} when available and from \citet{tj97} otherwise; when no \emph{R-I} value was available from either source, an assumed \emph{R-I} value of 2.00 was used, following \citet{chw04}.

\begin{figure}
\plotone{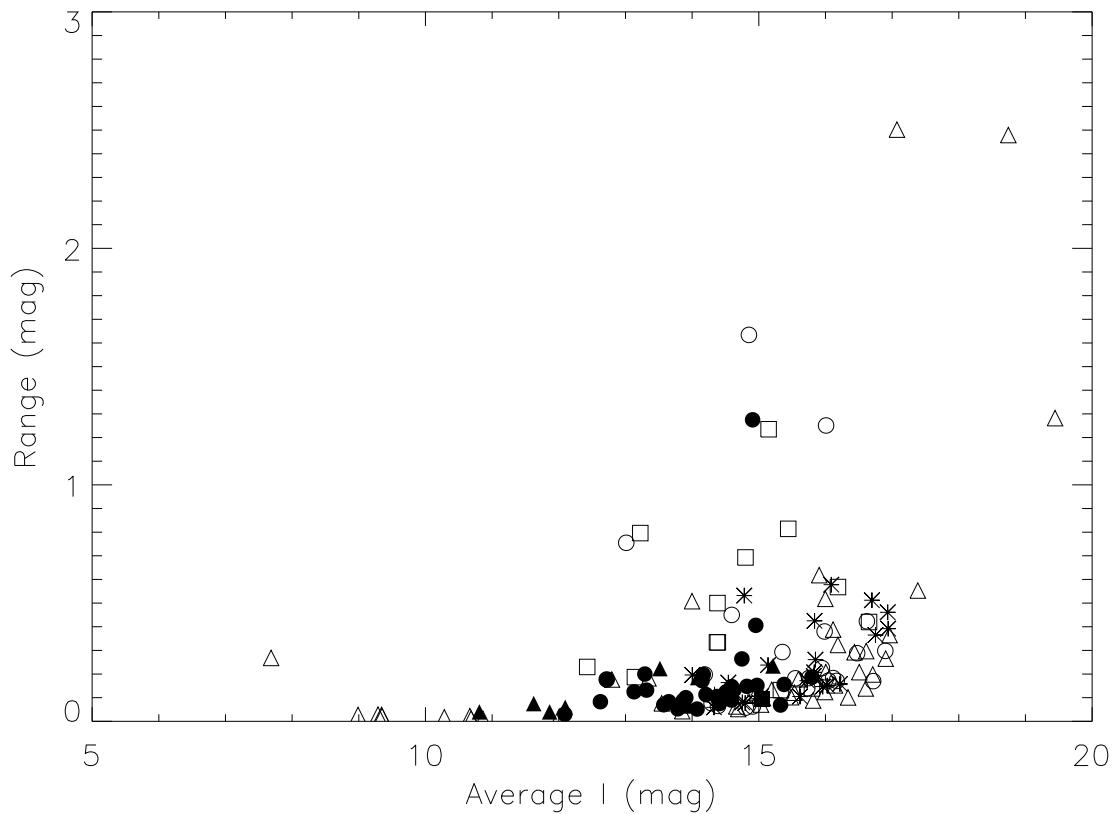}
\caption{Range v. Magnitude for
all the stars in our photometry sample, excepting those which are
photometrically contaminated by nearby stars. WTTS are plotted as
circles, CTTS are plotted as squares, and unclassified stars as
triangles. Plotting symbols for the periodic stars are filled, while
those of the non-periodic stars are empty. The absence of periods
among the highly variable stars (excepting star 73) is evident, as
is the lack of periods among both the brightest and the faintest
stars. The sample stars determined to be periodic by \citet{lit05} only
are plotted as asterisks. The majority of these newly-determined
periodic stars are at the faint end of the magnitude distribution,
supporting our belief that the lack of detected periods among stars
in our sample with $I>16$ is due primarily to signal-to-noise limitations of our photometry and not to the monitoring frequency.}
\label{range}
\end{figure}

Fig. \ref{range} shows the range of each star as a function of its
mean magnitude. As expected, there is a lower envelope that
increases in range with decreasing brightness due to random error.
At the bright end, our accuracy is about 0.005 mag, allowing us to
detect real variations as small as $\sim$0.02 mag for irregular
variables, or even smaller for periodic variables; where the signal
is concentrated into a narrow frequency band. For the fainter stars
in our sample, the errors grow sufficiently large that it is not
possible for us to detect periods, let alone real variability, even
when present. We return to a discussion of other features of Fig.
\ref{range} in the Results section.

\subsection{\emph{v} sin \emph{i} Measurements}

\subsubsection{Observations}

The \emph{v} sin \emph{i} observations were obtained in 2004 February
using the Hydra multiobject spectrograph on the 3.5 m WIYN
telescope\footnote{WIYN is a joint facility of the University of
Wisconsin, Yale University, and the National Optical Astronomy
Observatories.}\ at Kitt Peak National Observatory near Tucson,
Arizona. The Hydra instrument allows simultaneous spectroscopy of
$\sim$100 targets over a 1\degr\ field.  We used the same spectrograph
setup as for a previous study of rotation in the ONC
done by \citet{rhm01} [hereafter
RHM]: the Bench camera, red fiber cable, and 316@63.4 echelle
grating.  This setup yielded a resolution of R $\sim $21,500.  The
spectra cover $\sim$6240$-$6540 \AA\, with a central wavelength of
6400 \AA\ and 0.144 \AA\ per pixel dispersion.  The spectral resolution,
using a typical slit profile full-width half-maximum of 2 pixels, is
13.5 km~$s^{-1}$ at 6400 \AA.

The \emph{v} sin \emph{i} sample was chosen to include a mix of both
Classical T~Tauri stars (CTTS) and Weak T~Tauri stars (WTTS).  We also
wanted to observe roughly equal numbers of periodic stars and a
``control'' group of IC~348 members without measured periods.  (As we
note in what follows, several of the stars in the control group have
subsequently had their rotation periods measured.)  The general aim was to
compare the $v$~sin~$i$ distributions of these various samples with each
other, as was done in RHM.  The final set of target stars also
depended strongly on limitations on how the Hydra fibers could be
positioned.  Given all of these constraints, 37 target objects were
observed in two different fiber configurations, with 22 objects in the
first configuration and 15 in the second.  Both configurations
included 18 fibers positioned on the blank sky, to allow for
sky-subtraction in the spectral reduction process.

Five 1800-second exposures were taken in each configuration. Five
objects from the {\it Gliese Catalog of Nearby Stars} \citep{gj91}, with
spectral types roughly spanning the range of the target stars, were
observed for the RHM study with the same spectrograph setup, and were
used again here.  Spectra of these Gliese stars (which have Gliese
sequence numbers 15A, 75, 114, 144, and 411) were used as narrow-lined
templates in the cross-correlation process; their coordinates,
magnitudes and spectral types are given in RHM.

\subsubsection{Initial Reductions}

Preliminary data reductions were performed on both the template and
target star spectra using standard IRAF tasks such as ZEROCOMBINE,
CCDPROC, and FLATCOMBINE. The IRAF task DOHYDRA was then used on the
object frames to extract the spectra and perform flat-fielding, fiber
throughput correction, wavelength calibration, and sky subtraction.
The wavelength calibration was accomplished using observations of a
ThAr comparison lamp taken before and/or after each set of target star
integrations.  Sky fibers with unusually high signal due to object
contamination were eliminated from the sky subtraction process. The
remaining sky spectra were then averaged to create a combined sky
spectrum, using a sigma-clipping algorithm to eliminate cosmic rays.

After the DOHYDRA reductions were completed, the five individual
integrations of each target star were scaled and combined into a
single frame using the task SCOMBINE.  This produced a single high
signal-to-noise observation that was free of cosmic rays.  The
combined frames were then clipped so that the spectra contained only
the region between 6275 and 6525 \AA, to eliminate the lower
signal-to-noise regions at the ends of some of the images.  Finally,
the individual object spectra were continuum-normalized before the
cross-correlation process.

\subsubsection{Measuring Radial and Rotational Velocities}

Increasing rotational velocity causes the absorption lines in a
stellar spectrum to become broader and shallower.
When a broadened target star spectrum is cross-correlated against a
narrow-lined spectrum that otherwise contains similar spectral
features, the width of the Fourier cross-correlation peak depends upon
the amount of broadening present in the target spectrum.  The width of
this peak can therefore be used to derive the star's projected
rotational velocity. The position of the cross-correlation peak also
provides a measure of the radial velocity of the star with respect to
the template object, which can then be used to derive the target
star's heliocentric radial velocity.  We used the IRAF task FXCOR to
perform
cross-correlation of target star spectra against spectra from
narrow-lined template stars and derive radial and rotational
velocities.

The first step in deriving $v$~sin~$i$ is to establish the
relationship between the measured width of the cross-correlation peak
and $v$~sin~$i$; details of how this was done are given in RHM.
Briefly, the narrow-lined template spectra of the Gliese stars were
artificially ``spun-up'' to higher rotational velocities by convolving
them with a theoretical rotation profile (Gray 1992).  The broadened
spectra were then cross-correlated with the original versions.  The
FWHM values of the resultant peaks were measured and the relationship
between FWHM and $v$~sin~$i$ was quantified for each template star.

To estimate $v$~sin~$i$ of the target stars, each star's spectrum
was cross-correlated against the spectrum of the template star
having the closest spectral type.\footnote{The IC 348 spectral
types used in this study come from \citet{h98} when available or
\citet{l03} otherwise.}\ The FWHM of the cross-correlation peak
was measured and translated into a $v$~sin~$i$ value using the
previously-derived calibration curve for the template star.
Heliocentric radial velocities for  the target stars were
determined during the cross-correlation step using the measured
relative shift in the cross-correlation peak, the radial velocity
of the template star from \citet{gj91}, and the position, date,
and time of the observation.  The cross-correlation function can
also show structure that may indicate that a star is a
double-lined spectroscopic binary. In this case the presence of
two sets of spectral lines can cause the normally Gaussian-shaped
function peak to appear double-peaked, like two Gaussians
superimposed on one another. During the analysis, the
cross-correlation peaks were checked for evidence of possible
binary structure.

The limit to which \emph{v} sin \emph{i} can be measured depends in
large part on the size of the slit (fiber) image. The slit image width
of our instrumental setup, determined by measuring the FWHM of
emission-line profiles in the comparison lamp spectra, typically
ranged from 1.7 to 1.9 pixels in the middle of the spectrum to $\geq$2
pixels on the ends of the spectrum.  A 1.7-pixel spectral resolution
corresponds to a $v$~sin~$i$ of 11 km~$s^{-1}$, which we take as the {\it
approximate} lower limit on our \emph{v} sin \emph{i}
measurements. \emph{v} sin \emph{i} values in this paper that are
slightly above this limit should be treated with caution, as they
might be upper limits rather than actual velocity measurements.

\section{Results}

\subsection{Photometric Rotation Periods}
\subsubsection{Determination of Periods and Classification}
Periodic variations in the light curves of the sample stars were
searched for by calculating a periodogram for each star using the
method of \citet{s82} and the formulation of \citet{hb86}. This
method can yield multiple possible periods for a single star, some of which
are beat periods caused by the interference of the natural period
and the one-day sampling interval and others of which are false
alarms caused by non-periodic variations. In order to separate true
periods from aliases and false positives, the following
criteria were employed. First, a star's
power-spectrum (as generated by the periodogram function) must
display a peak with a normalized power greater than 7.0 and a false-alarm
probability (FAP), as defined by \citet{hb86}, of 2\% or less.
Second, the star must display one or more of the following
characteristics: (1) a well-defined light curve showing clear
periodic variation with small scatter, and/or (2) the highest periodogram
peak (exceeding a power of 7.0) recurring in multiple seasons.
Stars with periods between 0.9 and 1.1 days were
eliminated from consideration, as they could merely be an artifact of the one-day
sampling interval. With our sampling intervals we are simply not reliably sensitive 
to that frequency range. In cases when a star was determined to be
periodic in the majority of seasons, a search for the same period at a
slightly lower than 7.0 power was conducted and, when found, these periods were counted as
real detections in those seasons as long as they agreed to within 2\% with a period
had been identified in at least two other seasons.  

\begin{figure}
\plotone{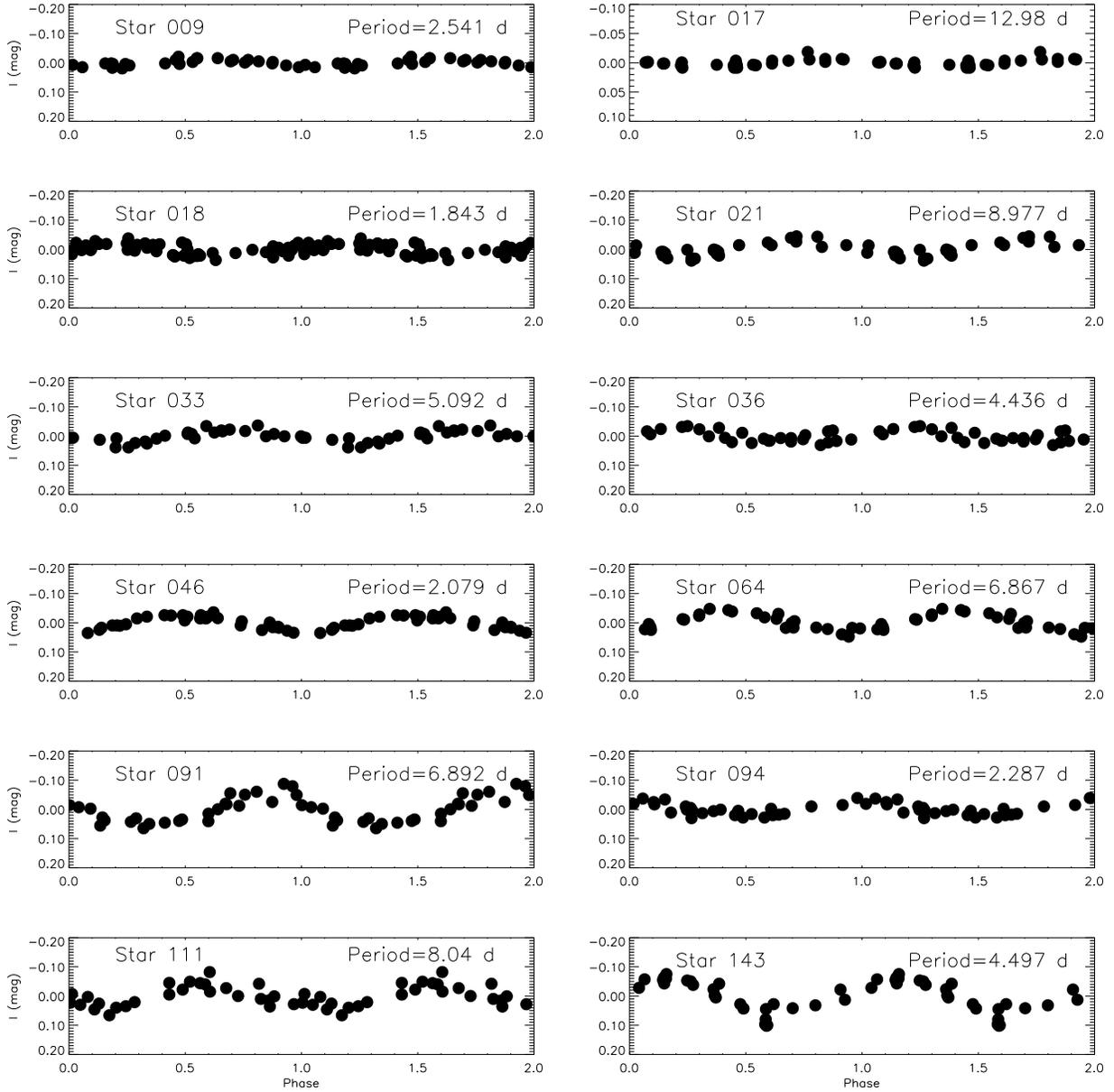}
\caption{Phased light curves for the newly detected periodic
stars. All are shown with 2004-2005 data except for star 46, which is
shown with 2002-03 data and star 18, which is shown with 1999-2000
data.}
\label{newperiods}
\end{figure}

Applying these
criteria, 40 of the 151 sample stars have now been determined to be periodic based solely on the VVO data set.
Twenty-eight of these stars were already identified by \citet{chw04} and 12 are new detections reported here. They are listed in Table \ref{data} and their phased light curves in one season are shown in Fig. \ref{newperiods}. Three of these had already been reported periodic by \citet{lit05} and two by \citet{kkb05}, although in one case the period reported by \citet{kkb05} is a beat period, not the fundamental period (see notes to Table \ref{data}).

Figure \ref{range} shows the relationship between range in magnitude
and average magnitude for all the stars in our sample. CTTS (squares) are defined here as K or M type stars with
$W_{H\alpha} > 11 $ \AA, while WTTS (circles) are defined as K
or M type stars with $W_{H\alpha} < 11 $ \AA. Spectral types and H$\alpha$ equivalent widths are taken from \citet{h98} or \citet{l03}.  It is evident that most of the stars found periodic
by us (solid symbols) tend to have fairly small ranges and be confined to the magnitude region $I \sim12-16$. It is also apparent that only one of the CTTS is periodic. The brightest cluster stars are of B to F spectral type and close to
the main sequence; as such we do not expect to find periodic
variations for them and do not. Periods are found for several of the relatively bright G-type stars.
The fact that we do no find periods among the stars fainter
than I = 16 is undoubtedly an artifact of the limitations imposed by the quality of our
photometry: \citet{lit05} were able to go deeper with better precision and discovered many periodic stars (asterisks on the figure) among stars near the faint end of our sample. 

On the other hand, the lack of detection of periods among the brighter stars (I = 14-16) with larger ranges, which are preferentially CTTS (see Fig. \ref{range}), is arguably not an artifact of our observing strategy. \citet{lit05}, for example, had a higher density of data points over a short period of time and they did not detect new periodic variations among these stars. They argue that they found CTTS  periodic variables that we missed because of their higher sampling frequency. However, we note that their new periodic CTTS are mostly fainter stars whose periodicity we probably missed because of the increased random errors in our photometry at those limits, not because of our sampling frequency. There remains a sample of relatively bright CTTS that have not revealed their periods, regardless of sampling frequency.

The lack of detectable periods
among highly-variable stars (with the exception of star 73, see
below) could be due either to the periodic variations being
overwhelmed beyond detection by large-amplitude irregular
variations or to a physical difference in stellar-spot presence or
configurations among highly-variable stars. The vast majority of 
periodic stars detected by us are WTTS. While sampling frequency
limitations could play some role, as argued above, such limitations were not responsible for our lack of detection of periodicity in the brighter CTTS.
\citet{lit05} found periods for a number (15) of fainter CTTS,
comprising about 30\% of their
periodic detections; these stars were too faint for us to find periods. Perhaps fainter CTTS reveal periodicity more readily than brighter ones. Another possibility is that the classification of CTTS
based on H$\alpha$ equivalent widths is somewhat more suspect for
faint stars, as the lower average signal can exaggerate the
prominence of their spectral lines. The average
range of CTTS in our sample is about 0.46 mag, compared to 0.24 mag for WTTS.

\subsubsection{Analysis of Periodic Behavior Over Time}

Assuming that T Tauri stars, like our own sun, demonstrate latitude drift of spots and 
differential rotation, one might expect to find changes in period and light-curve shape as stellar spots change in size and
shape and move to different stellar latitudes. However, we were
unable to find clear evidence of differential rotation in IC 348. In
general, our sample shows remarkable consistency with regard to
periodicity. Overall, the variation in the length of the period from
season to season was very small---in all but two cases, the periods
differed by less than 3\% over the entire 7-year time-span, with the
majority differing by less than 1.5\%. This change is below what we
consider to be significant, and is likely caused by slight changes
in spot configurations or phase changes. Of the remaining two stars,
one is a star which has only been found to be periodic in two
seasons, and thus could be the result of aliasing or coincidental
error, while the other varies by only 5.7\% over 7 seasons of monitoring, a result that is only marginally significant. The
consistency of period length in IC 348 stars seems to suggest that
either the rotation rate of T Tauri stars does not vary with
latitude, consistent with the rigid-rotator models of \citet{kr97},
or that T Tauri stellar-spots evolve or remain stable only at a
certain limited range of stellar latitudes. However, that we do observe large
variations in light-curve shape between different seasons seems to indicate a changing configuration of
spots over a range of latitudes. 

\begin{figure}
\plotone{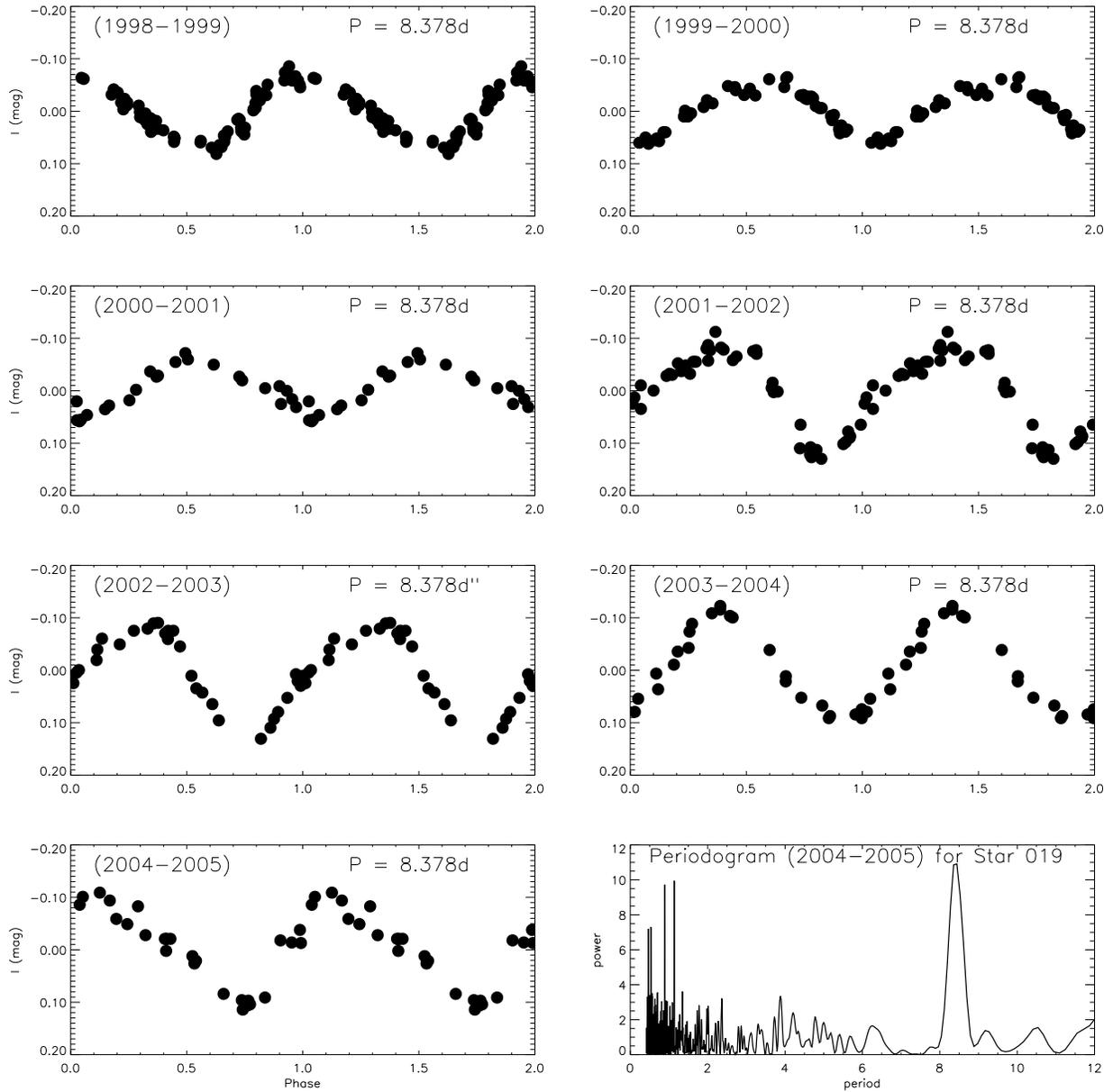}
\caption{Example light curve of a star (number 19)
with near constant periodic behavior throughout the 7-year observing
period. Changes in light curve shape are still apparent. The final
panel shows the periodogram (power vs. period) for 2004-2005.}
\label{goodmulti}
\end{figure}

\begin{figure}
\plotone{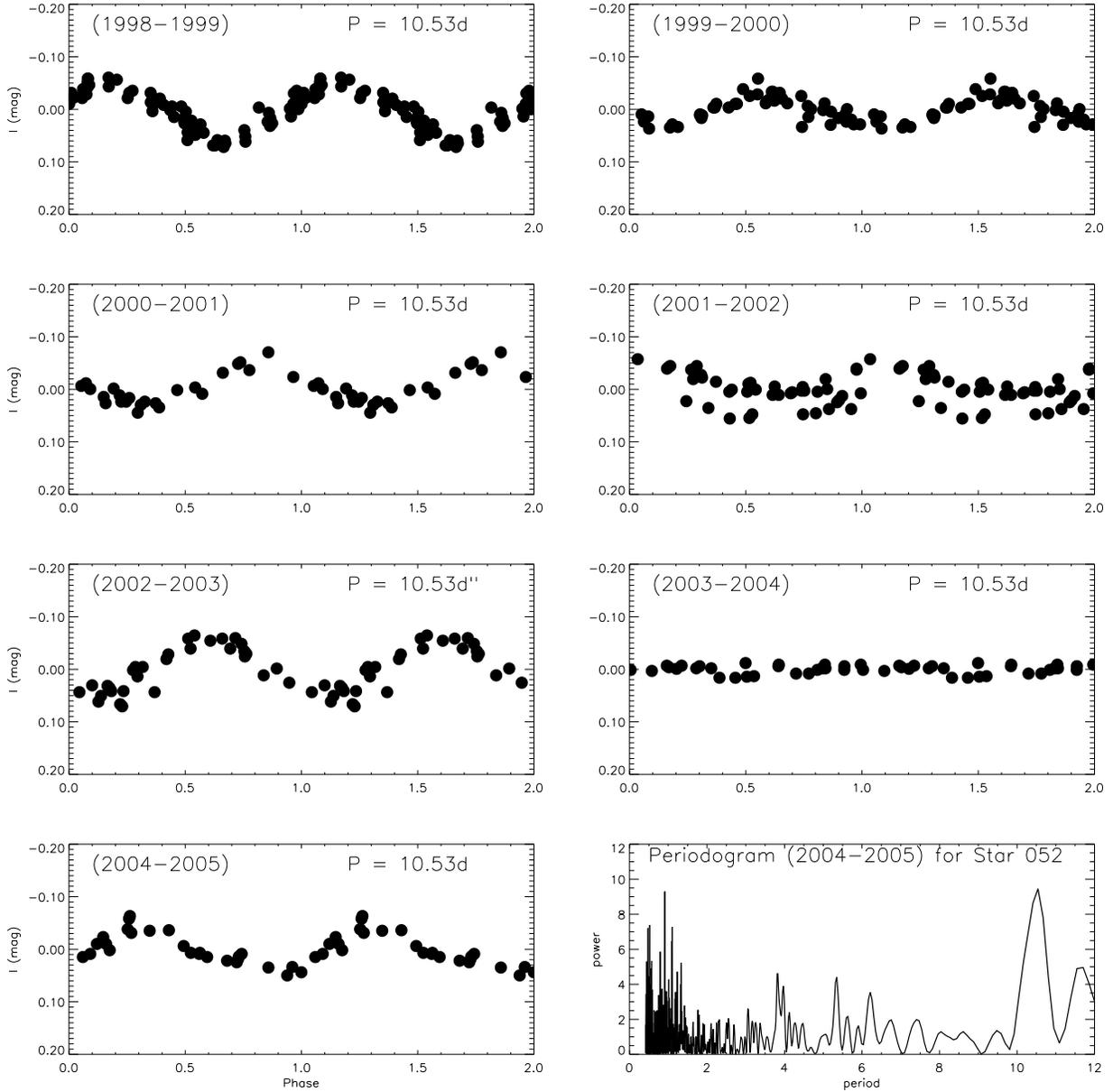}
\caption{Example light curve of a star (number 52)
for which the period was not detected in all years. For the
nonperiodic years (2001-2 and 2003-4), the data is phased using the
average detected period from the other years. The final panel shows
the periodogram (power vs. period) for 2004-2005.} \label{badmulti}
\end{figure}

Despite constancy of period length over the monitoring time-span,
all of the periodic stars do show slight changes in light-curve
shape and amplitude from year to year. We have looked for examples of
systematic cyclical changes, as one would expect
from stellar-spot cycles, and found none. Given that the duration of the
spot cycle on the Sun is 11.1 years, continued monitoring will clearly be
necessary to confirm or reject the possibility of similar stellar spot
cycles on T Tauri stars. Two examples of phased light-curves over
the 7 seasons are shown in Figures \ref{goodmulti} and \ref{badmulti}.
The light-curves were phased according to the periods detected
within each individual year, rather than using an average detected
period, as we found the individual-year periods to be less scattered in almost all
instances. Figure \ref{goodmulti} shows an example of a star which
was consistently periodic throughout the monitoring period, while
Figure \ref{badmulti} gives an example of a period which disappeared
and reappeared during the monitoring period.

The repeatability  of periodic behavior among IC 348 studies is also
confirmed by comparison between the periods detected through our
monitoring program and others. \citet{lit05} have
reported data on a total of 50 rotation periods in IC 348. Eight of
these are for stars too faint to be measured in our program, 11 were
not detectable as periodic in our data (likely due to the better
signal-to-noise ratio for the stars in the \citet{lit05} data: the
mean magnitude of these undetectable periodic stars is $I\sim15.4$,
while that of our periodic stars is $\sim13.8$), and 21 were
determined to be periodic by both studies. For the remaining 10
stars, we can identify agreeing periods in our data, but we had not reported them as periodic because they fail our identification
criteria, mostly by having too high a FAP. We can note these ten periods
as confirmations of those found by \citet{lit05}---periods which we
can identify as present in our data, but which do not show
sufficiently prominent periodic variations for us to identify them independently. They are listed in Table 2.
\citet{lit05} reported the same phenomenon for five of their stars, i.e. that we had reported them as periodic and that their data confirmed the periods but with a FAP that was too high to be regarded as definitive. Unfortunately, \citet{lit05} also took this to be evidence that our FAP criterion was more liberal than theirs. This is certainly not true.  The reality is that both groups have used criteria that are  conservative enough that real
periods have avoided detection as a result of the selection
restrictions. T Tauri stars vary in their amplitudes and light curve shapes from epoch to epoch resulting in natural variations in the power that one will obtain from the Lomb-Scargel periodogram. As a result, failure to detect a period in any given epoch at any given power level cannot be used to infer the accuracy of an FAP selection criterion.

\begin{figure}
\plotone{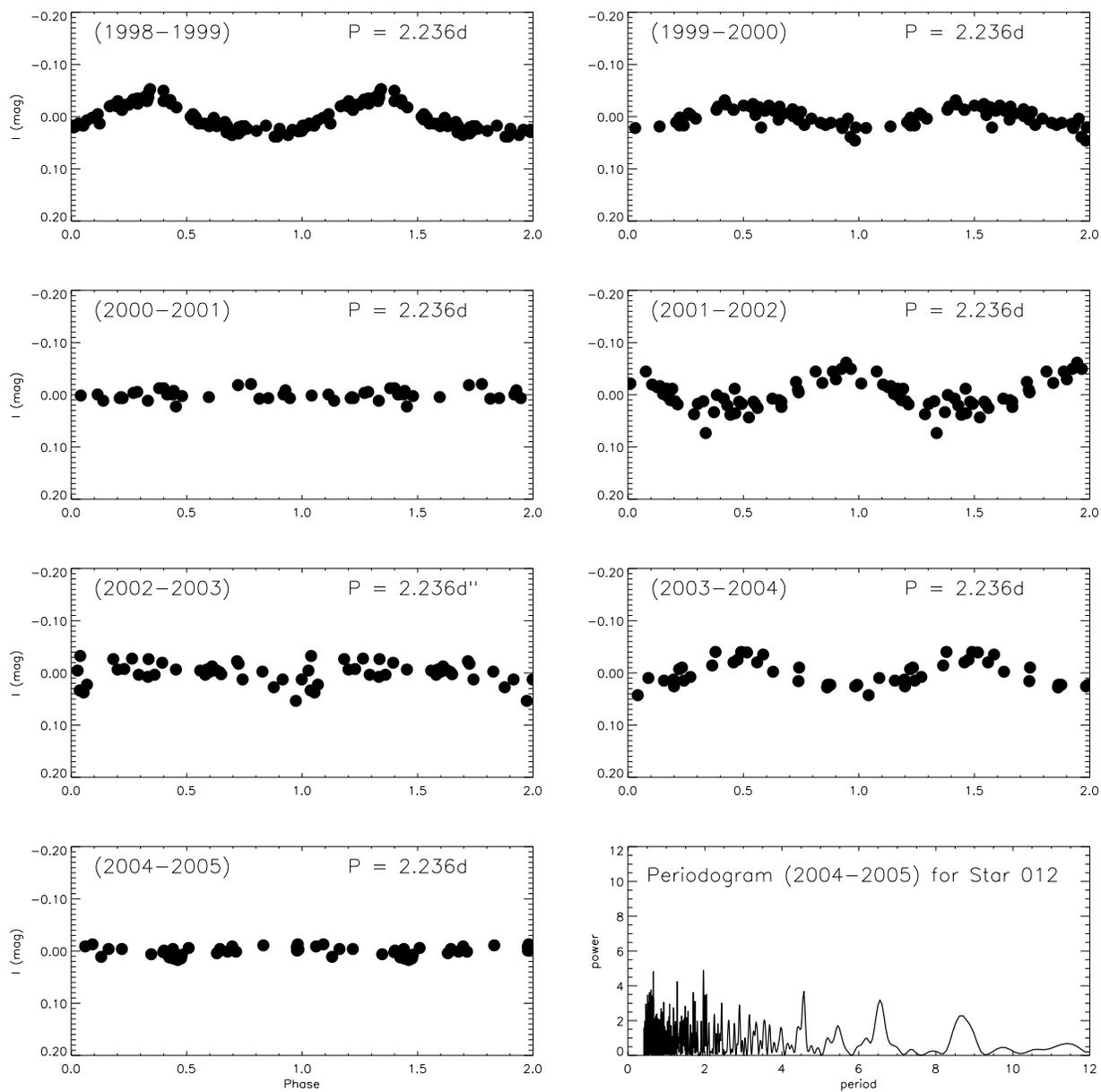}
\caption{Light curves of star 12, 
over 7 seasons phased with our identified period of $\sim2.2$ days. Clear
periodic structure can be seen in four seasons. The final panel
shows the 2004-05 periodogram for the star. \citet{lit05} report a period of 10.8 days for this star but the data phased with that period show no hint of periodicity in any season. We conclude that their determination is in error and that the actual period of the star is 2.24 days.} \label{star12}
\end{figure}

Overall, the agreement with \citet{lit05} is excellent; of the 21
independently found periodic stars the two studies had in common
(i.e. not counting the ``confirmations"), 19 of the detected
periods were in agreement to within the error of their period
determinations, about 0.1 days. This is gratifying confirmation of
the reliability of periodic variability studies. Of the remaining
two stars, one (our number 14) is the result of the beat
phenomenon, and we defer to the \citet{lit05} period of 1.63 days,
since their tightly-spaced monitoring is likely better suited to
distinguishing beat periods from true periods than our sparser
sampling. For the other star (our number 12), \citet{lit05}
reported a period of 10.8 days, while we detected a period of 2.24
days in 5 of our 7 monitoring years. Light curves of star 12 for
all seven seasons are shown phased for the 2.2 day period in Fig.
\ref{star12}. There is no evidence of a 10.8 day period in our
data for any of the sample years, while the 2.2 day period
produces a well-defined light curve in 4 of 7 seasons.
Additionally, our periodogram analysis shows no evidence for a
longer period. Given that our data extend over seven years and
include 274 nights of observing, while the \citet{lit05} study was
conducted over a much narrower time range (17 nights of data taken
over a span of 26 days), we conclude that $\sim2.24$ days is the
correct period for this star.

A second, even more recent study by \citet{kkb05} offers an additional opportunity for comparision--again, with excellent agreement. This five month study was able to confirm 17 of the periods identified in \citet{chw04} within 1\%; in addition, 18 new periods were detected. Eight of these stars were also studied in our monitoring program, and for three of them we have since found periods. One of these agrees with that found by \citet{kkb05}, while two (HMW 18 and HMW 143) do not, but follow the relationship  $1/p_{1} = 1 - 1/p_{2}$, marking one study's period as a beat period. As our data extend over a much longer period of time and offer stronger support for the periods detected in this study than in the \citet{kkb05} study, we conclude that our measurement is the more accurate one and list these stars as new detections. Of the remaining 5 detections by \citet{kkb05}, 3 can be confirmed in our data, though not with a low enough FAP to be treated as independent detections.

\subsubsection{Period Distributions}

\begin{figure}
\plotone{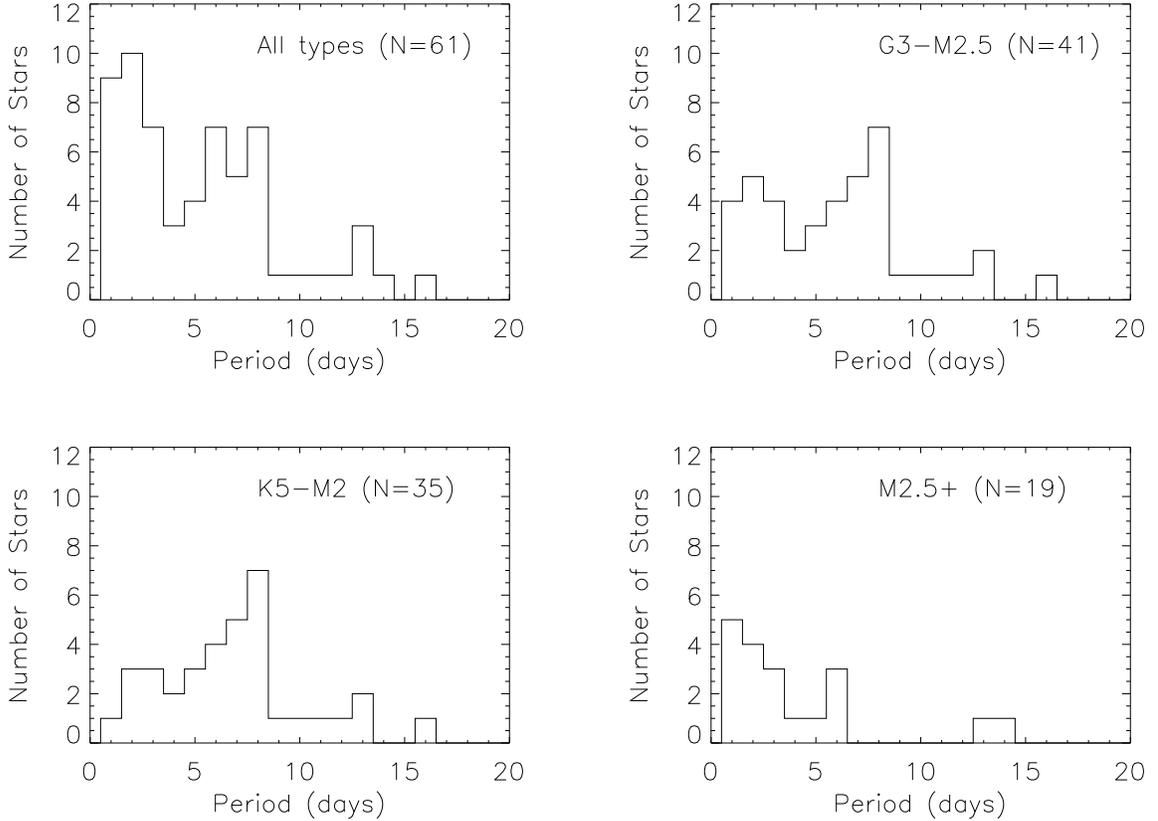}
\caption{ \emph{Top Left:} Distribution of
periods for all periodic stars in the sample, including the
additional stars found to be periodic by \citet{lit05}. \emph{Top
Right;} Distribution limited to stars of spectral type G3-M2.5. The
decrease in fast rotators when later-type stars are eliminated is
evident. \emph{Bottom Left:} Distribution limited to types K5-M2,
showing an additional decrease in fast rotators when early-type
stars are also eliminated.  \emph{Bottom Right:} Distribution for
the 19 stars with spectral types M2.5 and later, many of which are
too faint to be detected as periodic by our photometry, and thus
have periods measured only by \citet{lit05}. Note that there are no
stars in IC 348 with measured periods less than 1 day.} \label{hist_periods}
\end{figure}

Figure \ref{hist_periods} shows the distribution of periods in our
sample, as well as the effects of eliminating the earliest (pre-K5)
and latest (post-M2) stars from the distribution. The
main characteristics of the complete IC 348 sample distribution are an
absence of periods shorter than one day, a small number of periods
longer than 10 days, and a dip in the distribution of periods
between 3 and 5 days.  The first peak of the distribution (around 2
days) falls substantially when the distribution is limited to
stars earlier than spectral class M2.5, corresponding to mass $ > 0.25-0.4 M_{\sun}$, and nearly disappears when the
distribution is further limited to spectral types of K5 and later (mass $<$ $\sim$ 1.2 M$_{\sun}$),
demonstrating that nearly all of the earlier-type sample stars are
fast rotators (period $\leq 3$ days). This supports the inverse
relationship between rotation and mass for the more massive stars, found in \citet{chw04}. It is
also evident that most of the later-type (low mass) stars in the
sample also have relatively short periods ($<6$ days), similar to what is found in the
ONC and NGC 2264 \citep{hm05}. One sees clearly that, in this cluster, it is the intermediate mass stars (around 0.5-1 solar mass) that spin the slowest.

\begin{figure}
\plotone{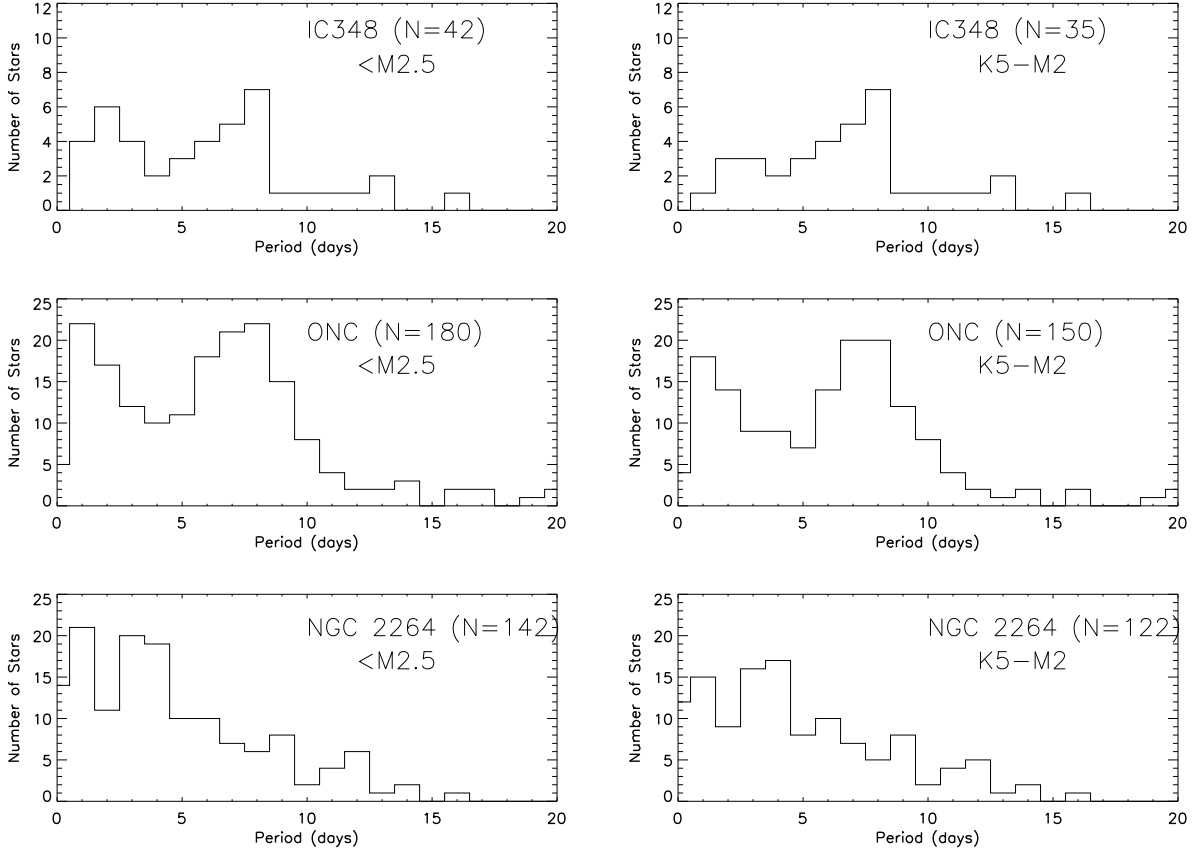}
\caption{Distribution of IC 348 periods
(\emph{top}) compared with those of the Orion Nebula Cluster (ONC -
\emph{center}) and NGC 2264 (\emph{bottom}) for stars of spectral
types earlier than M2.5 (left panels) and between K5 and M2 (right panels). In each case, the IC 348 and ONC distributions show roughly the same distribution and this is confirmed by a
Kolmogorov-Smirnov (K-S) test giving a 72\% (left) and 56\% (right) probability that the two
distributions are drawn from the same parent population. On the other hand, differences are seen between the IC 348 and NGC 2264 distributions which are significant at the 2 to 3 sigma level. A two-sided K-S test yields only a 3.3\% (left) or 0.4\% (right) probability that the data for the two different clusters are drawn from the same parent population.  Note that there are no
stars in IC 348 with measured periods less than 1 day.} \label{comphistearly}
\end{figure}

The distribution of
detected periods for stars with spectral types earlier than M2.5 in IC 348 is shown in the left hand panels of Fig. \ref{comphistearly} in comparison with like
distributions for the very young ($\sim1$ Myr) ONC and slightly
older (2-4 Myr) NGC 2264. The ONC data
is from \citet{hmb02}, classified based on the spectral types of
\citet{h97}, while the NGC 2264 data is from \citet{lam04}, classified based on the
spectral types of \citet{reb04}. The IC 348 and ONC distributions show
clearly similar double-peaked structures with peaks around 2 and 8
days; a Kolmogorov-Smirnov (K-S) Test on these distributions
indicates that they are likely ($\sim 72\%$ probability) drawn from
the same population. The NGC 2264 distribution, on the other hand,
is much more skewed towards fast rotators; the difference between
the IC 348 and NGC 2264 distributions is significant at a $2\sigma$
level, with a 3.28\% probability of being drawn from the same parent
population. 

The distributions are compared for stars of spectral
types K5-M2 in the right-hand panels of Fig. \ref{comphistearly}. Though there appear to be
some differences between the IC 348 and ONC distributions,
particularly a dearth of early rotators in IC 348, a K-S Test
shows that this difference is insignificant (57\% probability of
sharing the same parent population), largely due to the small size
of the IC 348 sample. The difference between the IC 348 and NGC 2264
distributions, however, is highly significant, with only a 0.4\%
chance of the two samples being derived from the same parent
population. The significance of these results for our understanding of the rotational evolution of solar-like stars is discussed in Section 4.

\subsubsection{Atypical Periodic and Irregular Variables}
\label{section:atypical variables}

\begin{figure}
\plotone{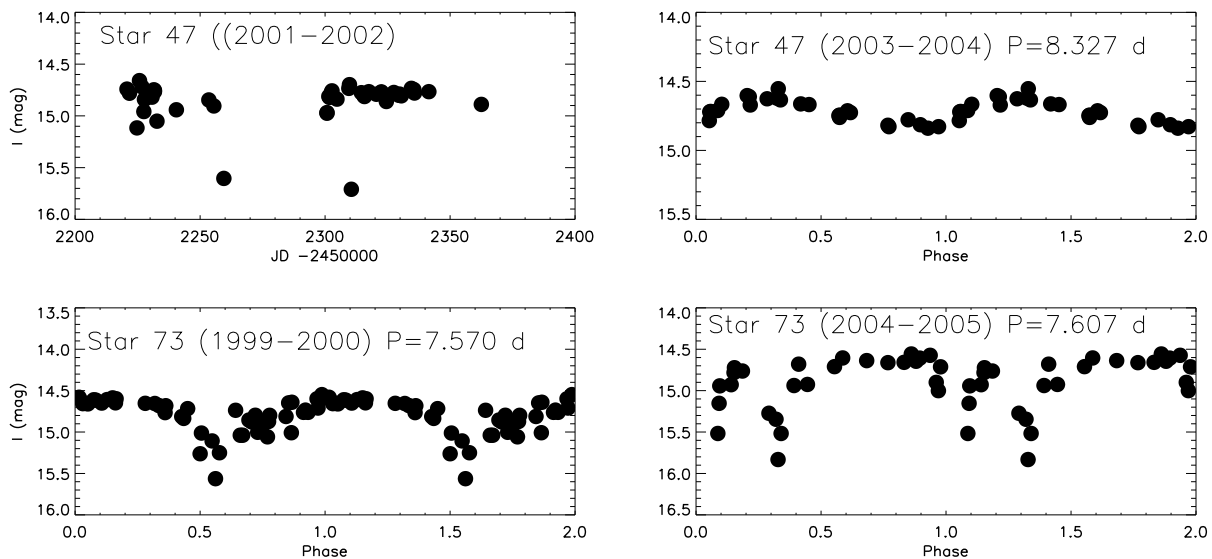}
\caption{Light Curve and Phase plots for
the two periodic and irregular stars in our sample. \emph{Top:} Star
47 shows highly irregular behavior in 2001-2002, characterized by
dips in brightness as large as $\sim1$ mag (\emph{left}), but shows
standard moderate-amplitude ($\sim0.25$) periodic behavior in
2004-2005 (\emph{right}.) \emph{Bottom:} Star 73 shows well-defined
but atypical periodic behavior in 1999-2000 (\emph{left}) but less
definitively periodic variations in 2004-2005 (\emph{right}.) In
both seasons the star decreases by about 1.3 magnitudes, a much
greater brightness change than any other periodic sample stars.}
\label{specialplot}
\end{figure}

In addition to periodic variability, many types of non-periodic
pre-main sequence variability have been identified over the years,
including FU and EX Orionis stars (FUors and EXors), as well as type II, and
type III (UXor) variables \citep{h77,h89,h94,hs99}. Sixteen
irregular variables in IC 348, mostly of type II, were identified and
discussed by \citet{chw04};  all these stars were monitored for an
additional two seasons in this study. Most of these stars continued
their irregular behavior, with one particularly remarkable
continuance observed for star 15 (to be documented in a forthcoming
paper.) \citet{chw04} identified one star (number 47) as both
periodic and irregular at different epochs. At that time, this K8 WTTS with I$\sim14.7$
had been seen as periodic in only one season (1998-1999). In the
next four seasons, it displayed no clear periodogram peak, but in
2000-01 and 2001-02 it displayed a light-curve with several deeper
minima, apparently fading by $\sim1$ mag in 1 day, then quickly
recovering to full brightness, a type of behavior more characteristic
of an UXor than a typical WTTS.  The new data show no more
irregularity, but clear periodic variations at a period within about
0.1 day of that detected in the first monitoring season. This star's odd
behavior is shown in the top two panels of Fig. \ref{specialplot}.
It certainly warrants further study in order to determine
whether its irregular behavior continues and deduce a possible
cause. Perhaps it is linked to the similarly odd star CB 34V \citep{thw03}.

A second periodic star which also displayed irregular variation
was found in this study: star 73, shown phase-plotted for two
years in the bottom two panels of Fig. \ref{specialplot} and also
noted as irregular in \citet{lit05}. Star 73, an M2 WTTS with an IR
excess of $\Delta(I-K)=0.64$ and a THICK disk classification based
on the slope of its infrared SED \citep{lm06}, has displayed
periods of duration 7.59 $\pm$ 0.08 days in four of the 7
monitoring seasons, and a period of 7.61 days at a slightly higher
FAP in the 7th season. These periodic variations, however, are on
a much larger amplitude scale ($\sim 1.3$ mag) than any other
periodic star, as shown in Fig. \ref{range}, where star 73 is the
clear periodic outlier. In order to explain a decrease in
brightness on this scale (by about a factor of three) using the
typical model of a rotating spotted surface, the star would have
to have either a very hot stellar spot or a cool stellar spot
taking up the vast majority of the surface area on one face.
Additionally, the light curve of star 73 takes a different shape
than any other periodic star, showing the sort of steep decline
immediately followed by a steep return to the original magnitude
typically associated with eclipsing binaries. This leads us to
believe that while this star does vary in a periodic way, the
cause of this variation may not be the rotation of a spotted
surface, as initially assumed. A spectrum of this star was
included in our $v \sin i$ study and showed no unusual structure;
additional study, particularly spectroscopic study, of this star
would help to shed light on the cause of its periodic and
non-periodic variation. As the star has an optically thick
circumstellar disk according to the slope of its infrared SED
\citep{lm06}, it is reasonable to suppose that, unlike the WTTS,
we are witnessing the effects of rotation of a rather stable
accretion hot spot. It is somewhat reminiscent of the odd behavior
exhibited by the CTTS AA Tau \citep{bga03}.

\subsection{\emph{v} sin \emph{i} Results}

Spectra of 37 stars were obtained for the purpose of measuring $v$~sin~$i$
and 30 of these produced reliable results.  For the remaining seven,
the spectra lacked sufficient signal to produce a clear
cross-correlation peak.  Half of the 30 stars with reliable
cross-correlations yielded $v$~sin~$i$ values less than or equal to our
estimated resolution limit of 11 km $s^{-1}$, so
the $v$~sin~$i$ values for these stars are upper limits.

Errors on our $v$~sin~$i$ values were calculated using the \emph{r}
parameter of \citet{td79}, which is a measure of the signal-to-noise
ratio of the peak of the cross-correlation function.
Tonry \& Davis showed that velocities measured with cross-correlation
should have errors that are proportional to $(1 + r)^{-1}$, and
\citet{h86} found in a study of Taurus-Auriga and Orion that $\pm v
\sin i/(1+r)$ was a good approximation of the 90\% confidence level of
their $v$~sin~$i$ measurements.  Here, as in RHM, we have adopted
$\pm v \sin i/(1+r)$ as an estimate of the $1\sigma$ errors
on our $v$~sin~$i$ values.

To explore whether this is a reasonable assumption, as well as to
investigate possible systematic errors, two of us (S.N. and K.R.)
independently executed the cross-correlation procedure to derive
$v$~sin~$i$ for the program stars.  When the FXCOR task is run, the
user interactively fits a Gaussian function to the cross-correlation
peak and decides exactly how the fit should be done (i.e., which data
points to include in the fit).  This has a direct impact on the FWHM
of the best-fit Gaussian function and therefore on the measured
$v$~sin~$i$.  For stars with $v \sin i$ above our 11 km~$s^{-1}$
resolution, the values measured by both of us agreed within the $\pm v
\sin i/(1+r)$ error estimate in all cases.  This suggests that the
$\pm v \sin i/(1+r)$ value is large enough to account for this
particular source of systematic measurement error.  Additional
possible sources of systematic error in our \emph{v} sin \emph{i}
measurement process include slight mismatches between the spectral
features of program stars and the cross-correlation template stars,
and a bias toward larger $v$~sin~$i$ due to line blending in the
stellar spectra.  Both of these issues were explored in depth by RHM
using WIYN/Hydra data taken with the identical spectrograph setup used
here, and with program stars of similar effective temperatures to the
IC~348 targets and the same set of template stars.  Based on that
analysis, we conclude that the $\pm v \sin i/(1+r)$ error estimate is
large enough to account for possible systematic uncertainties in the
current data set as well.

\begin{figure}
\plotone{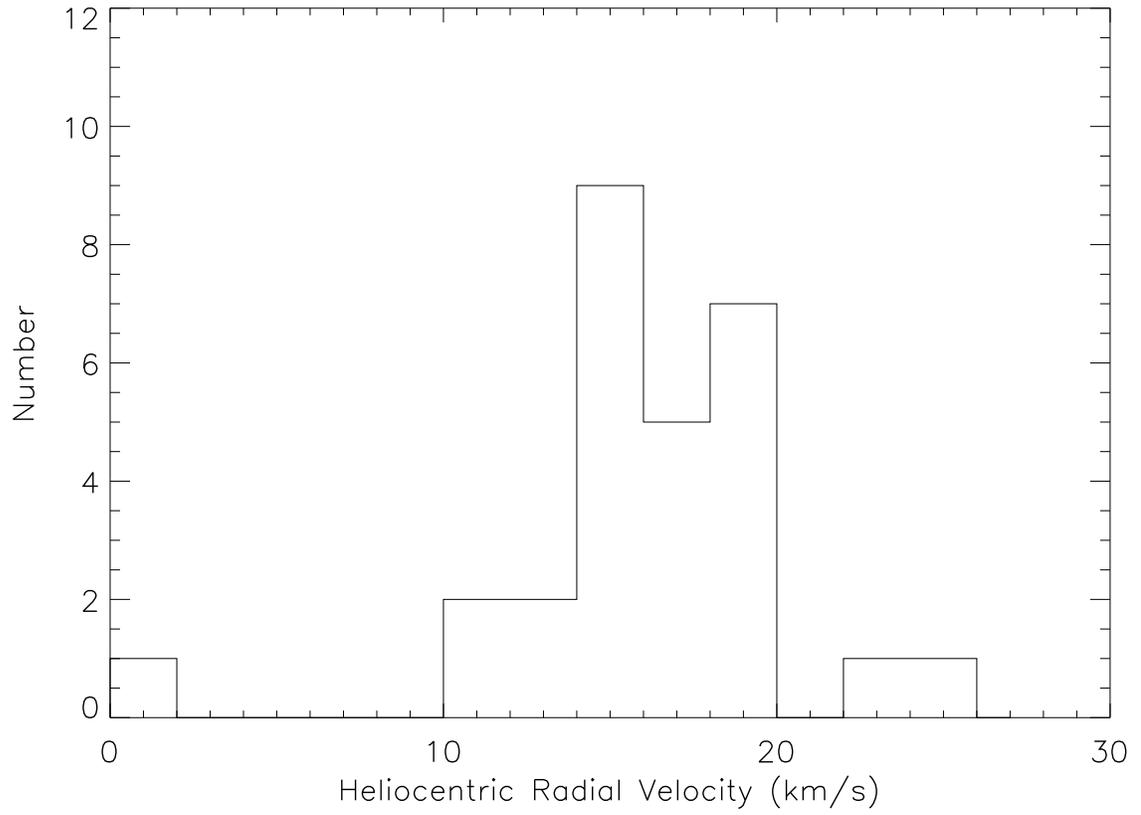}
 \caption{Histogram of the measured heliocentric radial
velocity values for cluster stars. Two possible SB2's with negative velocities are not shown. The mean and standard deviation calculated omitting those two stars are
16.5 km~s$^{-1}$ and 3.1 km~s$^{-1}$, respectively.} \label{hist_vhel}
\end{figure}

Two target stars (HMW~20 and HMW~45) showed structure in their
cross-correlation peaks that might indicate that they are double-lined
spectroscopic binaries (SB2s).  In both cases, the cross-correlation
peak had the expected Gaussian shape, but with a prominent ``bump'' or
second, lower-amplitude Gaussian peak on the side.  In addition to
$v$~sin~$i$, we also measured the target stars' radial velocities as
part of the cross-correlation process.  These radial velocity
measurements are, to our knowledge, the first for the
pre-main-sequence stars in IC~348. The distribution of these radial
velocities is shown in Figure \ref{hist_vhel}.  There are three
apparent outliers in the distribution: two of the stars (not plotted)
have negative radial velocities and are the possible SB2s identified
via structure in their cross-correlation functions; the other is
HMW~23.  The mean heliocentric radial velocity for our stellar sample
is 15.9$\pm$0.8~km~$s^{-1}$ if we exclude the two SB2 candidates and
16.5$\pm$0.6~km~$s^{-1}$ if we also exclude the outlier HMW~23.  These
values are close to the \citet{shss94} heliocentric radial velocity
estimate of approximately 12$-$15 km $s^{-1}$ for the supernova shell
in which IC~348 is embedded.

\begin{figure}
\plotone{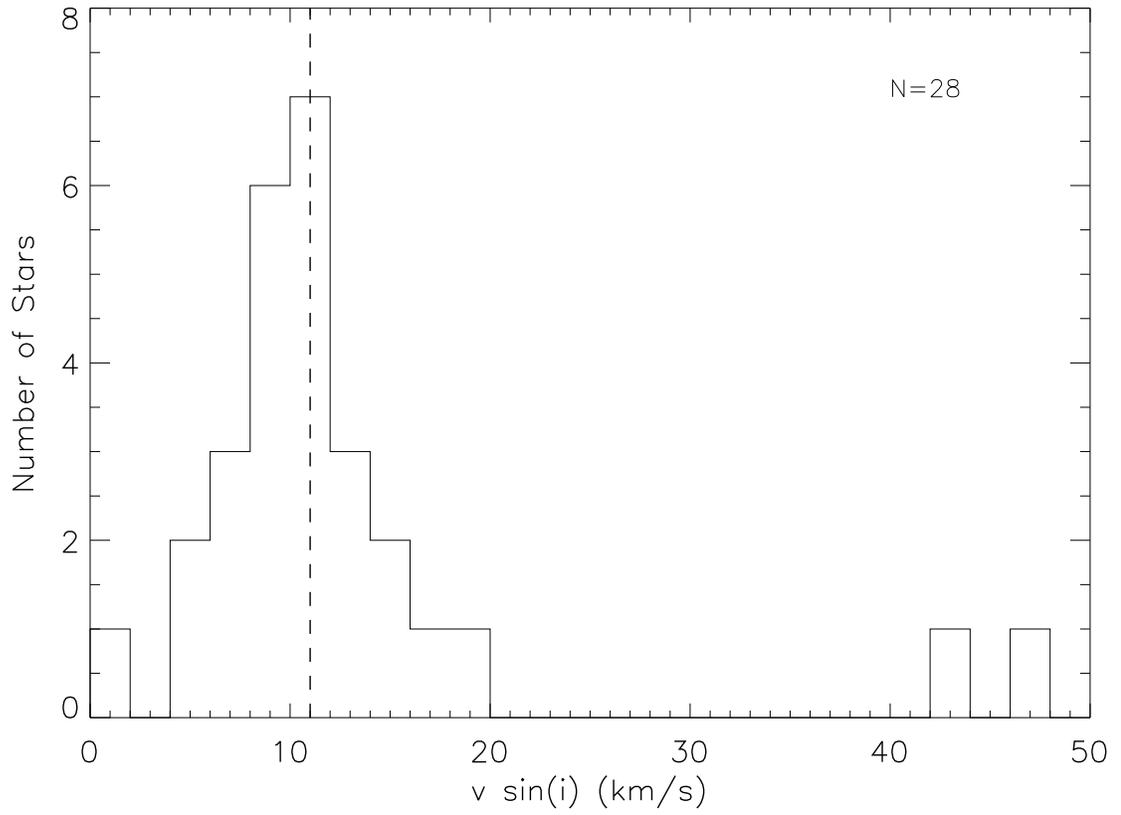} \label{hist_vsini}
\caption{Histogram of the distribution of \emph{v} sin
\emph{i} for the entire useable sample of 28 stars. The dashed line at 11 km/s indicates our estimated measurement limit and values below this have large percentage errors and should be considered only as upper limit measurements.}
\label{hist_vsini_per}
\end{figure}

The $v$~sin~$i$ measurements for the 30 stars with reliable
cross-correlation results are given in Table~\ref{vsiniresults}.  The
table lists sequence numbers for the stars (HMW and Luhman numbers),
heliocentric radial velocity, weighted mean $v$~sin~$i$ (calculated
using the independent $v$~sin~$i$ measurements done by S.N. and K.R.)
and the error on the mean ($\pm v \sin i/(1+r)$, where here $r$ is the
mean Tonry \& Davis $r$ value from the two measurements).  Stars with
$v$~sin~$i$ $<$ 11 km~s$^{-1}$ have ``$\leq$11.0'' in the $v$~sin~$i$
column.  Seven stars that were observed but that did not yield
$v$~sin~$i$ measurements (i.e., their cross-correlation results were
unreliable) are not listed in the table; they are HMW~31, 80, 88, 101,
110, 141, and 154.  A histogram showing the distribution of
$v$~sin~$i$ for the 30 stars is shown in Figure~\ref{hist_vsini_per}.

\subsubsection{Rotation Periods and $v \sin i$ Measurements}

\begin{figure}
\plotone{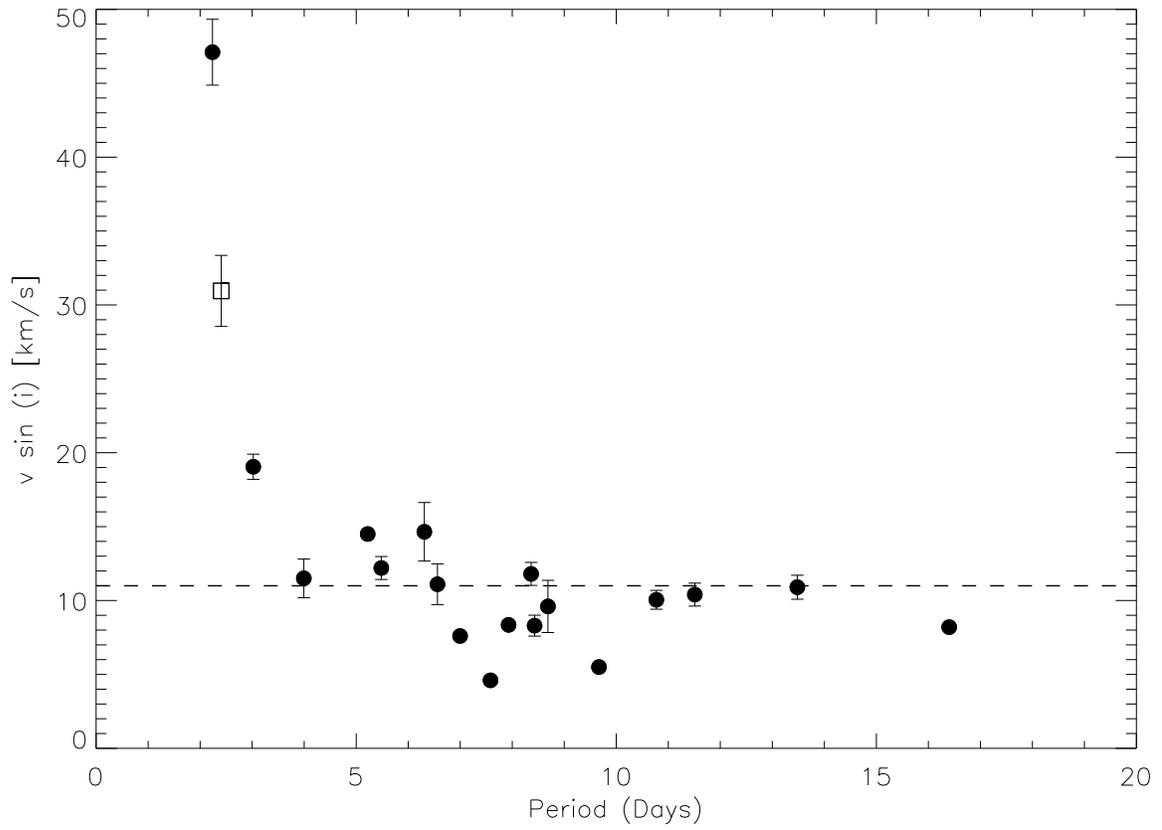}
\caption{Measured  \emph{v} sin \emph{i}
vs. period, showing the expected shape; a
possible double-lined spectroscopic binary (based on the line profile) is plotted as a square. The dashed line at 11 km/s indicates our estimated measurement limit and values below this have large percentage errors and should be considered only as upper limit measurements.} \label{peri_vsini}
\end{figure}

Assuming that the periodic behavior observed in the sample stars is
the result of the rotation of spotted surfaces, we would expect to see
a strong negative correlation between period and \emph{v} sin
\emph{i}'s, in that stars with shorter periods should display faster
rotational velocities. This relationship is shown in Fig.
\ref{peri_vsini}; although the stars with intermediate-length periods seem to
display slightly lower \emph{v} sin \emph{i} values than would be
expected, the difference in \emph{v} sin \emph{i} between the two
extremes of the period distribution is striking and offers strong
support for the spotted-surface assumption.  (A similar
conclusion was reached by both \citet{stassun99}, based on an
observed correlation between line width and angular velocity, and RHM, based on a highly significant period and $v$~sin~$i$ correlation,
for pre-main-sequence stars in Orion.)

In addition to a direct period--\emph{v} sin \emph{i} comparison, we
can use the measured rotation periods, in combination with radius
estimates, to derive equatorial rotation velocities from the equation
$v_{eq}=2\pi R/P$ and compare these values to the \emph{v} sin
\emph{i} measurements. The radii were derived from the relationship
$L=4\pi r^{2}\sigma T^{4}$, with the bolometric luminosities and
effective temperatures coming from \citet{l03}. For the stars for
which \citet{l03} did not report these values, the effective
temperatures were estimated based on spectral type using the scale of
\citet{ck79}, and the luminosities were calculated using the
formulation of \citet{h97} with a distance modulus of 7.5
\citep{h98}. The bolometric corrections for this calculation were
taken from \citet{b91} and \citet{bb88}, while the intrinsic stellar
colors are those of \citet{bb88}. A comparison of 
\emph{v} sin \emph{i} with the estimated rotational velocity shows the expected relationship. Despite the
relatively small number of data points, there is an obvious correlation between $v \sin i$ and $v_{eq}$, and nearly all the
stars respect the $v \sin i=v_{eq}$ limit within the estimated
error. This result again helps confirm our underlying assumptions
about T Tauri star rotation and periodic behavior. There is a slightly
anomalous concentration of stars above the $v \sin i=v_{eq}$ line
around $v \sin i \sim 10$, but as these stars all have $v \sin i$
measurements near or below our resolution limit, we conclude that this
is likely simply an artifact of the measurement process,
and that our $v \sin i$ results conform to previous assumptions about
pre--main-sequence rotation.

\begin{figure}
\plotone{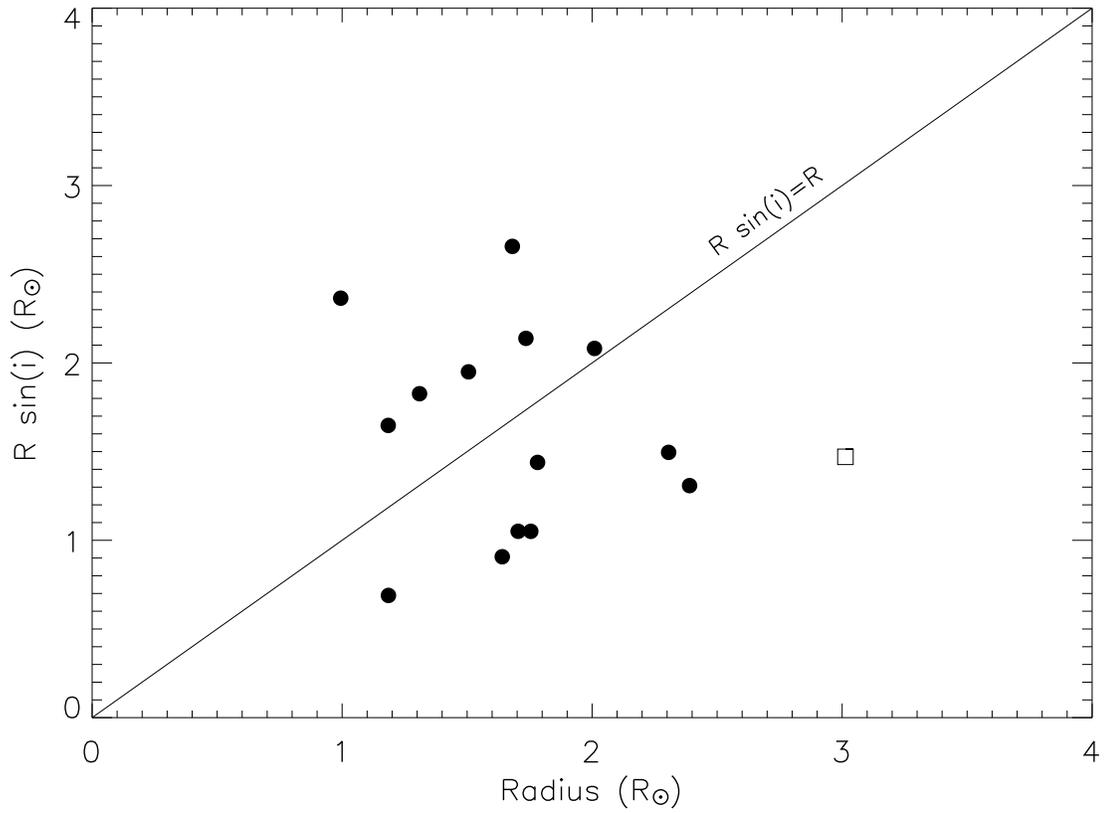} \caption{Measured \emph{R} sin \emph{i}
vs. radius, as calculated from luminosity and effective
temperature. The solid line is $ R \sin i=R$, and the possible
spectroscopic binary (SB) is plotted as a square. Note that
the SB candidate appears from its R value to be overluminous, which is
another piece of evidence suggesting it might be an SB.} \label{radrsini}
\end{figure}

The radii used in determining $v_{eq}$ are listed in column 6 of
Table \ref{vsiniresults}. They should be treated cautiously, as they
are highly sensitive to errors in luminosity and effective
temperature. In addition, column 7 lists the minimum radii ($R \sin
i$) derived from the relationship $ P v \sin i = 2\pi R \sin i$.
 Fig.  \ref{radrsini} plots the relationship between the two
quantities.  Although there are fairly large deviations between the
two for individual stars, likely due to the imprecision of the
luminosity and effective temperature measurements used in determining
the radii, overall the data points scatter around the line of
equality.  The mean calculated radius for the sample stars is
$R=1.75R_{\sun}$, while the mean $R \sin i$ for stars with $v \sin i$ measurements above the limit of 11 km/s is $1.62R_{\sun}$. We note that the mean radius (and, therefore, age) of the K5-M2 stars in IC 348 is nearly identical to that found for NGC 2264 (1.7 R$_{\sun}$) and significantly less than what is found for the ONC (2.1 R$_{\sun}$), as derived by \citet{hm05}.

\section{The Links Between Rotation Period, Photometric Variability and Circumstellar Disks}

The {\it Spitzer Infrared Telescope} is revolutionizing our understanding of star formation in many ways. Not the least of these is its ability to separate stars with infrared excess emission indicative of active accretion disks from those without it. The longer lever arm provided by this telescope's ability to probe further into the infrared with good signal-to-noise than has been possible heretofore has proven crucial to this endeavor. An excellent example of the value of such data for rotation studies is the work of \citet{reb06} who show clearly that in the ONC, rapidly rotating stars do not have disks, while more slowly rotating stars may or may not have them. Since the popular disk locking theory for slowing the rotation of stars \citep{k91} predicts, in its simplest form, some sort of correlation between rotation rate and the presence of a disk, we have investigated this in the present sample using Spitzer data from \citet{lm06}. They compute the slope of the spectral energy distribution (SED) in the infrared for cluster stars and use that to divide the stars into three classes, termed STAR, ANEMIC and THICK. The STAR class indicates no infrared excess and, therefore, no evidence for a circumstellar disk. ANEMIC and THICK refer to infrared SED slopes indicative of mild and substantial infrared excess, respectively. We have also investigated the possible link between large amplitude irregular variability and accretion disks. First, we discuss the periodic stars, then the irregular variables.

\subsection{Periodic Stars}

It has been known since the work of \citet{edw93} that there is a possible correlation between near-infrared excess (usually calculated as an excess in the I-K color index) and rotation. While the result has been challenged and the correlation is relatively weak and scattered for known reasons, it is nonetheless established at a very high level of statistical significance in the ONC \citep{hmb02}. Unfortunately, because of the natural scatter and nature of the correlation it requires a lot of stars to establish this in any cluster and there are relatively few stars available in IC 348. Furthermore, one must restrict such investigations to a fairly narrow mass range since rotation period is not just a function of infrared properties, it is also a strong function of mass. Here we follow previous work in the ONC and NGC 2264, as summarized recently by \citet{hm05}, and concentrate on the spectral class range of G to M2.

\begin{figure}
\plotone{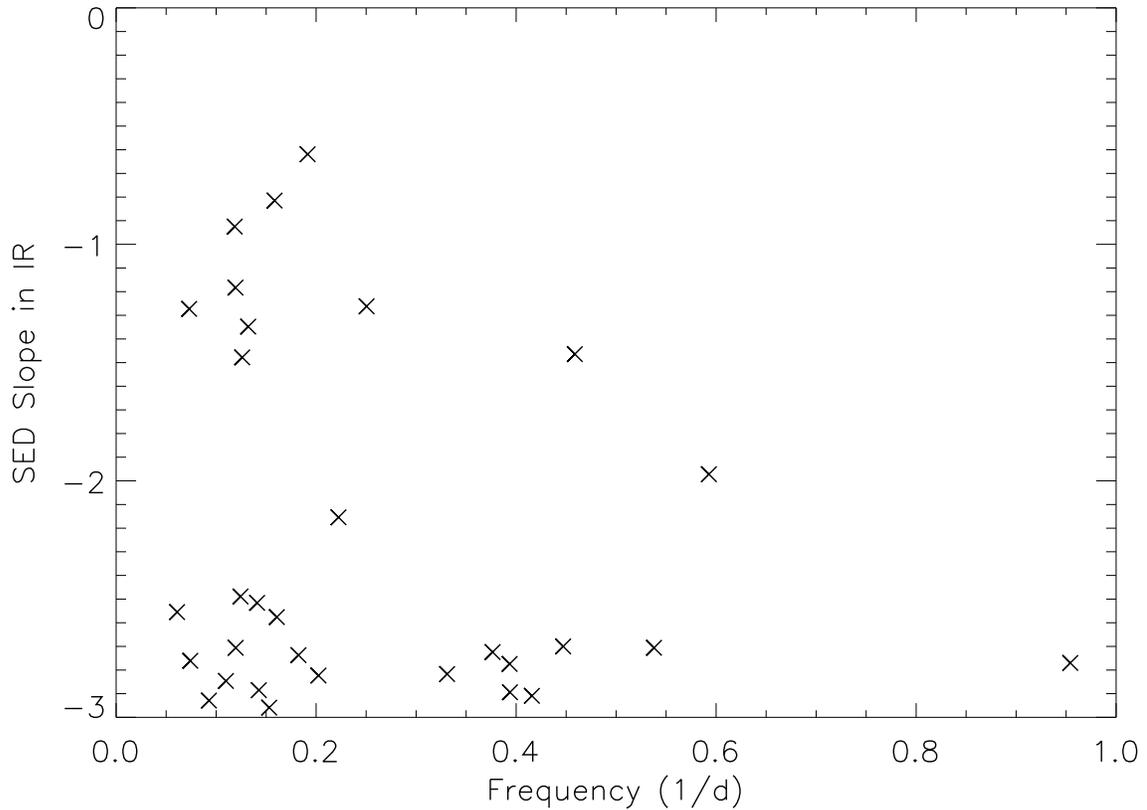} 
\caption{The slope of the mid-infrared
portion of the spectrum, as defined by Lada et al. (2006) on the
basis of their {\it Spitzer} data is shown as a function of
rotational frequency for G to M2 stars in IC 348. There is a weak
trend for rapid rotators (Period $<$ $\sim$3 days) to have large
negative slopes, indicative of little or no excess emission
characteristic of a disk. On the other hand, slow rotators have a
higher percentage of stars with shallower slopes, indicative of
infrared excesses attributed to circumstellar disks. While the
sample size is too small to be statistically significant, the
distribution is quite similar to what is found in the much larger
sample in the ONC.} \label{ir_freq}
\end{figure}

Figure \ref{ir_freq} shows the slope of the infrared portion of the SED, calculated by \citet{lm06} plotted against the rotation frequency (inverse of the period). On this plot, the slow rotators are to the left and rapid rotators to the right. There is a small natural gap at a frequency of 0.3, corresponding to a rotation period of about 3.3 days. This plot may be compared directly with the ONC as shown in the references cited above. While the number of stars is too small to be of statistical significance it is, nonetheless, notable that the general form of the distribution of points is quite similar to what is seen in the ONC. Namely, the rapid rotators tend to have less infrared excess (larger negative slopes), while the slower rotators have a mixture of stars with substantial excesses indicative of optically thick accretion disks and stars without such disks.

Because of the nature of this distribution and the small numbers involved it is not easy to quantify the statistical significance of any possible relationship between IR SED and rotation period. However, if we adopt the disk class assignments made by \citet{lm06}, the ratios of THICK:ANEMIC:STAR, for the rapid rotators (i.e period less than 3.3 days) are 1:1:8, while among the slow rotators they are 8:2:11. A one-sided Fisher's Exact Test yields a probability of only 14\% that rotation and IR SED are uncorrelated based on these numbers. Other statistical tests, including a Monte Carlo simulation, suggest the probability is only 7\% or less, but these rely on the sample size being large enough and that may not be the case here. Obviously these data alone would not constitute significant independent evidence for a disk-rotation link, but they do add some support to the picture because they are consistent with what is found in more populous clusters. In IC 348, as in the ONC and NGC 2264, slow rotators among the more massive stars are more likely to show infrared excesses indicative of optically thick disks than rapid rotators. Note that the interpretation of slow rotators without infrared excesses is that they are stars that have recently lost their disks and not yet had time to spin up. This is plausible because the time for a disk to disappear is probably the viscous accretion time scale, which is of order 10$^5$ years, while the spin-up time is an order of magnitude larger. We conclude that the infrared data are consistent with the view that disk-locking is operating and has operated in IC 348 to produce the large range in currently observed rotation rates, just as it has in the ONC (and NGC 2264).

\subsection{Irregular Variables}

\begin{figure}
\plotone{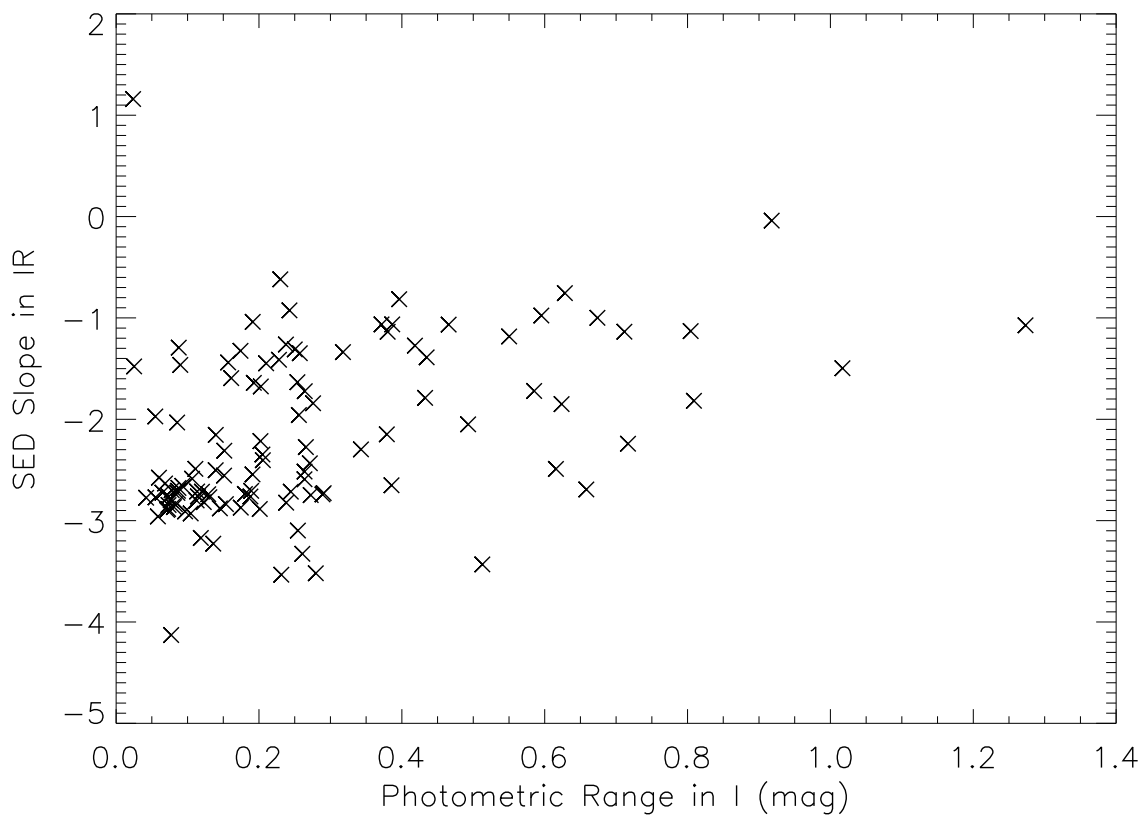}
 \caption{The slope of the mid-infrared
portion of the spectrum, as defined by Lada et al. (2006) on the
basis of their {\it Spitzer} data is shown as a function of
photometric range for G, K and M stars in IC 348. There is a clear
trend for large amplitude variables to have less negative slopes,
indicative of excess emission characteristic of a disk. Virtually
all stars with photometric ranges exceeding 0.4 mag in I have
evidence for disks. Large amplitude photometric variability is an
excellent predictor of IR excess in this sample. On the other
hand, IR excess emission by itself does not mean a star will
necessarily be a large amplitude photometric variable. Absence of
IR excess emission does mean that a star in this cluster is very
unlikely to be photometrically active at a level exceeding 0.3 mag
in I.} \label{ir_range}
\end{figure}

Fig. \ref{ir_range} displays the relationship between a star's infrared properties and its variability range (which, as shown in Fig. \ref{range},  is largest for the irregular variables.) Evidently, there is a strong and highly statistically significant correlation between our measured photometric range for a star and its SED slope in the infrared from \citet{lm06}. The sense of the correlation is that stars with evidence for circumstellar disks tend to have larger photometric ranges, exactly what we expect based on the canonical picture of CTTS variability as arising from unsteady accretion luminosity. Of the 21 stars in our sample with I magnitude ranges exceeding 0.4 mag, 19 have SED slopes indicative of ANEMIC or THICK disks. Large amplitude photometric variability is an excellent predictor of the presence of a circumstellar disk.

On the other hand, the absence of large amplitude variability does not necessarily preclude the presence of a disk. Many low amplitude variables are classified as ANEMIC or THICK by \citet{lm06}. Presumably these are stars that were, for one reason or another, accreting at a steadier rate or less actively accreting during the observing epoch. It will be interesting to learn on what time scale they change this behavior and why, but that will require additional monitoring.
These findings confirm the basic expectation about variability among T
Tauri stars that CTTS will vary by larger amounts and more irregularly due
to unsteady accretion, presumably from a circumstellar disk, while WTTS variability is predominantly from the rotational modulation of a spotted photosphere.

\section{Discussion: A Role for Environment in Establishing Initial Rotation Rates of Stars?}

Finally, we consider the origin of the fact that the period
distribution in IC 348 of the G-M2 stars is similar to that of the
ONC but significantly different from NGC 2264. The first thought
would be that IC 348 and the ONC share a similar age. However,
this is probably not true based on our best estimate of the radii
of the pre-main sequence stars. Concentrating on the narrow
spectral range K5-M2, and using the luminosities and effective
temperature given by \cite{l03, l05} we find that the average $R$ 
of these stars is $1.67 \pm 0.17 R_{\sun}$ . This may be
compared to values of 2.1 and 1.7  R$_{\sun}$ for the ONC and NGC
2264, respectively \citep{hm05}. These estimates, of course, do depend on
the assumed distances to the clusters involved, and the reader is referred to
\citet{h06} for a discussion of the distance to IC 348 and its error. Most of the debate
has been over whether the cluster could be closer than the adopted 300 pc, which
would make the stars smaller and older. With the commonly adopted distances, IC 348 is very close in
age, as judged by mean radius of pre-main sequence stars, to NGC
2264 and significantly older than the ONC. These radii differences
correspond to an age factor of about 2: IC 348 and NGC 2264 are
about twice as old as the ONC. For a representative ONC age of 1
Myr, that makes IC 348 about 2 Myr old.

One would have expected that, all things being equal, the rotation period distribution of IC 348 would match that of NGC 2264, not the ONC. Clearly, something besides mass and age is responsible for influencing the rotation of stars in different clusters. An obvious possibility is the lifetime of disks, since it is disks that evidently control rotation during the pre-main sequence stage. To understand the observations one would require that disks last for a longer time in IC 348 than they do in either of the other clusters mentioned. This seems quite reasonable since the other clusters have O stars within them, while the earliest type star in IC 348 is of spectral class B5. The lack of ionizing photons means that the environment of the pre-main sequence stars is less hostile to disks than in the more massive clusters. This, in turn may mean that the disks could survive longer in the smaller cluster.

Our suggestion would be, therefore, that the rotation of stars
will be influenced by the cluster environment in which they form.
In particular, if it is a dense cluster with O stars then the
disks will not last long and the stars will spin up earlier and
more of them will end up as generally faster rotators on the ZAMS.
In sparse clusters or associations, where disk lifetimes may be
longer, one would expect more disk braking and, therefore, a
larger proportion of slower rotators when they reach the ZAMS.
This suggestion is consistent with the observations (although not
the precise interpretation) of B-type stars in young clusters and
the field, as discussed recently by \citet{sw05}. Obviously it
will require many more observations of many more clusters and
associations to ascertain whether this proposal has any merit. 

\section{Summary and Conclusions}

This study has confirmed several beliefs about T Tauri star
variability and rotation based on other clusters, identified some
interesting stars worthy of continued observation and provided a
challenge to the simplest picture of rotational evolution of low mass
stars. In particular, the new data confirm that WTTS variability
arises primarily from the rotation of a spotted photosphere and that
the large amplitude (greater than 0.4-0.5 mag in I) irregular
variability of CTTS is linked to the presence of an accretion
disk. This study also verifies the accuracy, reliability and
remarkable consistency of periods of pre-main sequence stars over time
scales of many years and confirms the link between rotation and
photometric period through comparison with $v \sin i$ data. 

The most unusual variable in IC 348 is HMW 15 and it is the
subject of a separate, forthcoming contribution. Also notable is
HMW 73, the only star to show large amplitude periodic behavior.
This may be the result of rotation of a stable accretion hot spot
or more unusual variable behavior.

Finally, we have shown here that the rotation period distribution
of stars in the 0.4-1.2 solar mass range in IC 348 matches that of
the ONC but contains a higher proportion of slow rotators than is
seen in NGC 2264. Since the age of IC 348 is closer to NGC 2264
than the ONC, this suggests that age and mass alone are not
sufficient to predict the rotation properties of stars in an
extremely young cluster. We suggest that environment may also be a
factor, as \citet{sw05} have also proposed based on their v sin i
study of h and $\chi$ Persei. If disk-locking controls rotational
evolution for stars in this mass and age range then the longer
accretion disks can persist, the longer the stars can continue to
resist spin-up due to contraction with conservation of angular
momentum. Perhaps in the less harsh environment of IC 348, whose
earliest type star is of B5 and not a prodigious source of
ultraviolet photons, the disks can persist longer than in NGC
2264. If so, the stars may continue to reflect an ONC-like
distribution for times scales of order 2 My.

Continued study of this, and similar clusters, will be needed to
ascertain whether this environment argument has merit. Because the
rotation period distribution is so broad (and mass dependent) for
young clusters, one requires a large number of stars to determine
it and to distinguish significant differences among clusters.
 Obviously, it will continue to be of
interest to monitor the variations of peculiar young stars such as
HMW 15 and HMW 73 since they undoubtedly have much to tell us
about the processes of accretion and disk evolution in solar-like
stars.

\acknowledgments It is a pleasure to thank the Wesleyan students
who carried out the observations. S. N.'s work was supported by the National Science
Foundation under Grant No. AST-0353997 to Wesleyan University,
supporting the Keck Northeast Astronomy Consortium. K. R. was
supported by an NSF Astronomy \& Astrophysics Postdoctoral
Fellowship award under AST-0302095. This material is based upon work supported by the National Aeronautics and Space Administration under Grant NNG05GO47G issued through the Origins of Solar Systems Program to W. H.

\clearpage

\begin{deluxetable}{ccc}
\tabletypesize{\scriptsize} \tablecaption{Newly Detected Periodic
Stars at VVO \label{data}} \tablewidth{0pt} \tablehead { \colhead{Star
Number}   & \colhead{Period (Days)} & \colhead{Notes}}

\startdata
 9 & 2.54 & \emph{a}  \\
 17 & 12.98 \\
 18 & 1.84 & \emph{b}\\
 21 & 8.98 & \emph{c} \\
 33 & 5.09 \\
 36 & 4.44 \\
 46 & 2.08 & \emph{c} \\
 64 & 6.69 & \emph{c} \\
 91 & 6.89 \\
 94 & 2.29 \\
 111 & 8.04 \\
 143 & 4.50 & \emph{d} \\
 \enddata
\tablenotetext{a}{Similar period detected in \citet{kkb05}.}
\tablenotetext{b}B{eat period reported by \citet{kkb05}.}
 \tablenotetext{c}{Similar period detected in \citet{lit05}.}
 \tablenotetext{d}{Could be the result of periodic variation in either star 143 or 144, which are photometrically indistinguishable.}
\end{deluxetable}

\begin{deluxetable}{cccc}

\tabletypesize{\scriptsize} \tablecaption{Stars Detected as Periodic Both at VVO and by Littlefair et al. (2005). \label{comp}} \tablewidth{0pt} \tablehead { \colhead{Star
Number}   & \colhead{Period (Days)} & \colhead{Littlefair et al. Period (Days)} & \colhead{Notes}}

\startdata
12 & 2.24 & 10.81 & \emph{a} \\
14 & 2.54 & 1.63 & \emph{b} \\
16 & 5.22 & 5.21 & \\
21 & 8.98 & 9.11 & \\
26 & 3.02 & 3.01 & \\
27 & 2.67 & 2.65 & \\
29 & 8.61 & 7.93 & \emph{*} \\
30 & 7.57 & 7.55 & \\
31 & 3.09 & 3.08 & \\
32 & 4.94 & 4.90 & \\
34 & 6.35 & 8.43 & \emph{*,c}\\
37 & 13.37 & 14.00 & \emph{*} \\
39 & 9.67 & 10.16 & \\
41 & 7.00 & 7.12 & \\
45 & 2.41 & 2.39 & \\
46 & 2.08 & 2.09 & \\
50 & 5.48 & 5.36 & \\
51 & 11.51 & 12.43 & \\
52 & 10.77 & 11.22 & \\
60 & 7.09 & 6.96 & \\
64 & 6.87 & 6.84 & \\
69 & 3.26 & 3.27 & \\
74 & 2.27 & 2.25 & \emph{*} \\
81 & 3.99 & 3.98 & \\
98 & 3.51 & 3.58 & \emph{*}\\
107 & 2.14 & 1.90 & \emph{*,d}\\
122 & 1.54 & 1.54 & \emph{*} \\
133 & 6.35 & 6.31 & \emph{*}\\
134 & 8.69 & 8.56 & \\
135 & 7.47 & 7.48 & \emph{*}\\
142 & 2.98 & 1.68 & \emph{*,d}\\

 \enddata
 \tablenotetext{*}{An asterisk in the `Notes' column indicates that
for this star, we were able to confirm the period detected by
Littlefair et al (2005)---i.e., a similar period was present in
our data, though it did not meet our criteria for independent
detection.} \tablenotetext{a}{Clear periodic variations at
$\sim2.24$ days were observed in several seasons of data for this
star, whereas we found no evidence of a 10.8 day period in any of
the seven seasons. Given that our data span a much larger time
range, and would thus be well suited to detecting any
long-duration ($\sim10$ day) period that existed, we conclude that
2.24 days is the more accurate measurement.}
 \tablenotetext{b}{In this case, we expect that our period may be a beat period, and defer to the shorter period identified by Littlefair et al (2005.)}
\tablenotetext{c} {Although there is a fairly large discrepancy
between the two periods, the 6.5 day period shows up strongly in
our data, albeit below our normal detection criteria.}
 \tablenotetext{d}{Here we have detected a beat period for the Littlefair period, indirectly confirming the period identified in that study.}
\end{deluxetable}

\begin{deluxetable}{rrrrrrlc}
\tabletypesize{\scriptsize} \tablecaption{$v \sin i$ Results
\label{vsiniresults}} \tablewidth{0pt} \tablehead { \colhead{HMW \#} & \colhead{Luhman \#} & \colhead{$v_{\rm helio}$} & \colhead{$v \sin
i$} & \colhead{Period} & \colhead{$R$} & \colhead{$R~\sin i$} & \colhead{Notes}\\
\colhead{} & \colhead{} & \colhead{(km~s$^{-1}$)}
&\colhead{(km~s$^{-1}$)} & \colhead{(days)} & \colhead{($R_{\sun}$)} &
\colhead{($R_{\sun}$)} & \colhead{}}
\startdata
12     & L29  &  15.0$\pm$4.3 &  47.1$\pm$3.3 & 2.24 & 2.01 & 2.08 & \\
16     & L36  &  23.2$\pm$2.0 &  14.5$\pm$0.9 & 5.22 & 2.31 & 1.50 & \\
20     & L5   &  -4.9$\pm$4.0 &  26.5$\pm$2.8 & ... & ... & ... & \emph{a} \\
22     & ...  &  17.5$\pm$3.8 &  11.8$\pm$1.4 & ... & ... & ... & \\
23     & L37  &   0.2$\pm$2.5 & $\leq$11.0 &    ... & ... & ... & \\
26     & ...  &  17.1$\pm$2.4 &  19.1$\pm$1.4 & 3.02 & ... & 1.14 & \\
29     & L32  &  18.1$\pm$2.8 & $\leq$11.0 &    7.93 & 2.39 & 1.31 & \\
34     & ...  &  14.0$\pm$4.4 & $\leq$11.0 &    8.43 & ... & 1.38 & \\
39     & ...  &  15.0$\pm$3.3 & $\leq$11.0 &    9.67 & 1.75 & 1.05 & \\
40     & L125 &  15.7$\pm$3.4 &  11.8$\pm$1.2 & 8.36 & 1.51 & 1.95 & \\
41     & ...  &  19.2$\pm$1.9 & $\leq$11.0 &    7.00 & 1.70 & 1.05 & \\
42     & L86  &  13.3$\pm$5.3 &  11.1$\pm$2.0 & 6.56 & 1.78 & 1.44 & \\
44     & L65  &  18.7$\pm$1.8 & $\leq$11.0 &    16.40 & 1.68 & 2.66 & \\
45     & L23  &  -9.4$\pm$5.3 &  30.9$\pm$3.8 & 2.41 & 3.01 & 1.47 & \emph{a} \\
48     & L113 &  18.1$\pm$4.5 & $\leq$11.0 &    ... & ... & ... & \\
50     & ...  &  17.0$\pm$3.1 &  12.2$\pm$1.2 & 5.48 & ... & 1.32 & \\
51     & ...  &  16.4$\pm$4.2 & $\leq$11.0 &    11.51 & 0.99 & 2.37 & \\
52     & ...  &  18.0$\pm$2.9 & $\leq$11.0 &    10.77 & 1.74 & 2.14 & \\
53     & ...  &  14.3$\pm$4.0 & $\leq$11.0 &    13.48 & ... & 2.90 & \\
56     & L75  &  14.1$\pm$5.7 &  13.9$\pm$2.3 & ... & ... & ... & \\
59     & L41  &  13.9$\pm$4.8 & $\leq$11.0 &    ... & ... & ... & \\
65     & L61  &  24.9$\pm$5.7 &  17.3$\pm$2.7 & ... & ... & ... & \\
73     & ...  &  14.4$\pm$4.3 & $\leq$11.0 &    7.58 & 1.19 & 0.69 & \emph{b} \\
75     & L40  &  18.9$\pm$3.8 &  13.7$\pm$1.5 & ... & ... & ... & \\
78     & L45  &  18.8$\pm$5.2 &  42.9$\pm$3.6 & ... & ... & ... & \\
81     & L91  &  11.5$\pm$5.3 &  11.5$\pm$1.8 & 3.99 & 1.64 & 0.91 & \\
133    & L60  &  15.8$\pm$6.7 &  14.7$\pm$2.8 & 6.31 & 1.31 & 1.83 & \emph{c} \\
134    & ...  &  11.1$\pm$7.9 & $\leq$11.0 &    8.69 & 1.18 & 1.65 & \\
152    & L67  &  16.9$\pm$2.9 & $\leq$11.0 &    ... & ... & ... & \\
153    & L63  &  14.7$\pm$4.7 & $\leq$11.0 &    ... & ... & ... & \\
\enddata
\tablenotetext{a}{Structure in the cross-correlation function
indicates that this star is a possible double-lined spectroscopic
binary.}
 \tablenotetext{b}{There is some uncertainty about the periodicity of
   this star --- see Section~\ref{section:atypical variables}.}
 \tablenotetext{c}{Period from \citet{lit05}.}

\end{deluxetable}

\clearpage

\clearpage

\clearpage

\clearpage

\clearpage

\clearpage

\clearpage

\clearpage

\clearpage

\clearpage

\clearpage

\end{document}